\documentclass[%
 reprint,
superscriptaddress,
nofootinbib,
 amsmath,amssymb,
 aps,
]{revtex4-1}

\usepackage{graphicx}
\usepackage{dcolumn}
\usepackage{bm}
\usepackage{xcolor}
\usepackage{hyperref}

\usepackage{subfigure}
\usepackage{supertabular}
\usepackage{widetable}

\newcommand{\dcc}{LIGO--P1900029}
\newcommand{\pol}{\psi_{\rm pol}}
\newcommand{\fstar}{f_\star}
\newcommand{\Tcoh}{T_{\rm coh}}
\newcommand{\Tobs}{T_{\rm obs}}
\newcommand{\fprec}{f_{\rm prec}}

\graphicspath{{./fig/}}

\begin{document}

\preprint{APS/123-QED}

\title{Tracking continuous gravitational waves from a neutron star at once and twice \\the spin frequency with a hidden Markov model}

\author{Ling Sun}
\email[]{lssun@caltech.edu}
\affiliation{LIGO Laboratory, California Institute of Technology, Pasadena, California 91125, USA}

\author{Andrew Melatos}
\affiliation{OzGrav, University of Melbourne, Parkville, Victoria 3010, Australia}

\author{Paul D. Lasky}
\affiliation{OzGrav, School of Physics and Astronomy, Monash University, Clayton, Victoria 3800, Australia}

\date{\today}

\begin{abstract}


Searches for continuous gravitational waves from rapidly spinning neutron stars normally assume that the star rotates about one of its principal axes of moment of inertia, and hence the gravitational radiation emits only at twice the spin frequency of the star, $2\fstar$. The superfluid interior of a star pinned to the crust along an axis nonaligned with any of its principal axes allows the star to emit gravitational waves at both $\fstar$ and $2\fstar$, even without free precession, a phenomenon not clearly observed in known pulsars. The dual-harmonic emission mechanism motivates searches combining the two frequency components of a signal to improve signal-to-noise ratio. We describe an economical, semicoherent, dual-harmonic search method, combined with a maximum likelihood coherent matched filter, $\mathcal{F}$-statistic, and improved from an existing hidden Markov model (HMM) tracking scheme to track two frequency components simultaneously. We validate the method and demonstrate its performance through Monte Carlo simulations. We find that for sources emitting gravitational waves at both $\fstar$ and $2\fstar$, the rate of correctly recovering synthetic signals (i.e., detection efficiency), at a given false alarm probability, can be improved by $\sim 10\%$--70\% by tracking two frequencies simultaneously compared to tracking a single component only. For sources emitting at $2\fstar$ only, dual-harmonic tracking only leads to minor sensitivity loss, producing $\lesssim 10\%$ lower detection efficiency than tracking a single component. In directed continuous-wave searches where $\fstar$ is unknown and hence the full frequency band is searched, the computationally efficient HMM tracking algorithm provides an option of conducting both the dual-harmonic search and the conventional single frequency tracking to obtain optimal sensitivity, with a typical run time of $\sim 10^3$\,core-hr for one year's observation.

\end{abstract}

\maketitle


\section{Introduction}

Continuous waves, produced by rapidly rotating neutron stars, including isolated stars and the ones in binary systems, are persistent, quasimonochromatic gravitational-wave signals detectable by ground-based interferometers such as the Laser Interferometer Gravitational Wave Observatory (LIGO) and the Virgo detector \cite{LIGO2014,Virgo2014,Riles2017}. Depending on the generation mechanisms, the neutron stars are expected to emit gravitational radiation at specific multiples of the star's spin frequency, $f_\star$ \cite{Lasky2015,Riles2017}. A persistent thermoelastic or magnetic mass quadrupole produces emission at $f_\star$ and/or 2$f_\star$ \cite{ushomirsky00, Johnson-McDaniel2013, Cutler2002, Mastrano2011, Lasky2013}. An r-mode current quadrupole produces emission roughly at 4$f_\star$/3 \cite{Owen1998, Heyl2002, Arras2002, bondarescu09}. A current quadrupole due to nonaxisymmetric circulation in the superfluid interior pinned to the crust emits at $f_\star$ \cite{Peralta2006, VanEysden2008, Bennett2010, Melatos2015}. The emission spectrum of a triaxial star may contain peaks at more frequencies, depending on the source orientation.

In most of the continuous-wave searches to date, an optimal scenario of a perpendicular rotor spinning about one of its principal axes of moment of inertia is considered, and hence the gravitational waves are only emitted at $2\fstar$ \cite{Riles2017}. More generally, when the star's rotation axis and its principal axis of the moment of inertia do not coincide, spanning an angle $\theta$, a nonaligned rotor freely precesses, and emits gravitational waves mainly at $\fstar$ and $2\fstar$, and weakly at a number of other frequencies \cite{Zimmermann1979, Zimmermann1980,Jones2002,VanDenBroeck2005,Lasky2013}. However, there is no clear observational evidence of free precession in the population of known pulsars (although see Refs.~\cite{Stairs2000,Jones2001,Chung2008,Kerr2016,Jones2017,Ashton2017}), which is one of the reasons that a perpendicular rotor is generally considered in most continuous-wave searches. 

\citet{Jones2010} considered a model that a neutron star contains a superfluid interior pinned to the solid crust along an axis that is not aligned with any of the star's principal axes of moment of inertia. The pinned superfluid inside the crust adds extra angular momentum to the system, such that the star's total angular moment vector coincides with its rotation axis. Hence the star can steadily rotate without free precession, even though none of its crustal principal axes is aligned with its rotation axis. In this case, the gravitational-wave emission is at both $\fstar$ and $2\fstar$. Unlike a triaxial precessing star, the gravitational-wave spectrum of a triaxial star with pinned superfluid interior does not involve weak emission at frequencies in addition to $\fstar$ and $2\fstar$. In a special case, when the star is a nonperpendicular biaxial rotor, the signal waveform proposed by Ref.~\cite{Jones2010}, composed of two frequency components, is identical to that from a biaxial precessing star \cite{Zimmermann1979}. 

The pinned superfluid model has been adopted in \emph{targeted} searches for known pulsars \cite{Bejger2014,Pitkin2015}, using ephemerides measured electromagnetically from absolute pulse numbering. In the data collected by the initial LIGO in the fifth science run (S5), searches were carried out for 43 known pulsars at both $\fstar$ and $2\fstar$, and the first upper limits on the gravitational-wave strain amplitude at two frequencies were set \cite{Pitkin2015}.
Recently, searches have been conducted for 222 known pulsars at both $\fstar$ and $2\fstar$, using the most sensitive data from the first two observing runs of Advanced LIGO (O1 and O2), and new upper limits have been placed on the gravitational-wave strain amplitude, mass quadrupole moment, and fiducial ellipticity \cite{O2-target}.
However, in \emph{directed} continuous-wave searches, search methods scan templates without guidance from an electromagnetically measured ephemeris due to the lack of timing data, although the sky position of the source can be known precisely from photon astronomy. Hence \emph{directed} searches are generally more expensive than \emph{targeted} searches. All of the existing \emph{directed} searches assume the $2\fstar$ only emission for simplicity \cite{Riles2017,Sun2016,Aasi2015-snr,O1-snr}. 

In this paper, we introduce an approach based on a hidden Markov model (HMM) \cite{Quinn2001}, which provides an economical solution to track both $\fstar$ and $2\fstar$ simultaneously in a stack-slide-based semicoherent directed search. A HMM tracks unobservable, time-varying signal parameters (hidden states) by relating them to the observed data through a likelihood statistic in a Markov chain. The Viterbi algorithm \cite{Viterbi1967} provides a computationally efficient HMM solution, finding the most probable sequence of hidden states. The technique was applied to a search for continuous waves from the most luminous low-mass x-ray binary, Scorpius X-1, in the Advanced LIGO O1 run \cite{Suvorova2016,ScoX1ViterbiO1},
and a search for long-transient signals from a postmerger remnant of the binary neutron star merger GW170817 in O2 \cite{LVC:2018pmr,Sun:2018hmm}. The technique is also proposed as an economical alternative to other stack-slide-based semicoherent methods in young neutron star searches \cite{Sun2018}. Here we extend the algorithm to dual-harmonic tracking, which takes into consideration the model of a nonperpendicular biaxial rotor in addition to the conventional perpendicular biaxial rotor model in directed continuous-wave searches, without introducing much additional computing cost. We demonstrate the sensitivity improvement through systematic simulations.

The structure of the paper is as follows. In Sec.~\ref{sec:sig_model}, we review the signal model of gravitational waves from a neutron star emitting at both $\fstar$ and $2\fstar$. We briefly describe a frequency domain maximum likelihood matched filter $\mathcal{F}$-statistic in Sec.~\ref{sec:F-stat}. In Sec.~\ref{sec:hmm}, we formulate the dual-harmonic HMM tracking scheme, implement a semicoherent search strategy, and discuss the analytic path probability distribution. In Sec.~\ref{sec:simulation}, we quantify the sensitivity improvement of tracking two frequency components compared to tracking a single component only through Monte Carlo simulations. The computing cost and potential applications of the method are discussed in Sec.~\ref{sec:discussion}. A summary of the conclusions is given in Sec.~\ref{sec:conclusion}.

\section{Signal model}
\label{sec:sig_model}
In this section, we review the phase of the continuous wave signal observed at the detector on Earth (\ref{sec:phase}), and describe three signal models: a perpendicular biaxial rotor (\ref{sec:triaxial-aligned}), a nonperpendicular biaxial rotor (\ref{sec:biaxial-case}), and a triaxial nonaligned rotor (\ref{sec:triaxial-non-aligned}). 
 
\subsection{Signal phase}
\label{sec:phase}
Taking into consideration the Doppler modulation of the observed signal frequency due to the motion of both the Earth and the neutron star with respect to the solar system barycentre (SSB), the signal phase observed at the detector is given by \cite{Jaranowski1998}
\begin{equation}
\label{eqn:phase}
\Phi(t) = \Phi_0 + 2\pi \sum_{k=0}^{s} \frac{\fstar^{(k)}t^{k+1}}{(k+1)!}+ \frac{2\pi}{c}\hat{n}\cdot\vec{r}(t)\sum_{k=0}^{s}\frac{\fstar^{(k)}t^k}{k!},
\end{equation}
where $\Phi_0$ is the initial phase at reference time $t=0$, $\fstar^{(k)}$ is the $k$-th time derivative of the spin frequency of the neutron star at $t=0$, $\hat{n}$ is the unit vector pointing from the SSB to the star, $\vec{r}(t)$ is the position vector of the detector relative to the SSB, and $c$ is the speed of light. 

\subsection{Perpendicular biaxial rotor}
\label{sec:triaxial-aligned}
Let $I_1$, $I_2$, and $I_3$ be the three principal moments of inertia of the star, the simplest model is a perpendicular biaxial rotor with $I_1=I_2$, equivalent to a triaxial rotor spinning about one of the principal axes.
The dimensionless amplitude of the gravitational-wave signal is
\begin{equation}
\label{eqn:h0}
h_0 = \frac{16 \pi^2 \fstar^2(I_3-I_1)}{r},
\end{equation}
where $r$ is the distance from the Earth to the star.
The gravitational-wave emission is at $2\fstar$ only, with plus and cross polarized amplitudes
\begin{eqnarray}
\label{eqn:h2p_classic} h_{2+} &=& \frac{1}{2}h_0(1+\cos^2\iota)\cos 2\Phi ,\\
\label{eqn:h2c_classic} h_{2\times} &=& h_0\cos \iota  \sin 2 \Phi,
\end{eqnarray}
where $\iota$ is the inclination angle of the source.
The signal can be written in the form
\begin{equation}
\label{eqn:sig_form_simplest}
h(t)=\sum_{m=1}^{4}\mathcal{A}_{m} h_{m}(t),
\end{equation}
where $\mathcal{A}_m$ denotes the amplitudes, depending on $h_0$, $\Phi_0$, $\iota$, and the wave polarization angle $\pol$. They are associated with the linearly independent components
\begin{eqnarray}
\label{eqn:h1} h_1(t) &=& a(t) \cos \Phi(t), \\
h_2(t) &=& b(t) \cos \Phi(t), \\
h_3(t) &=& a(t) \sin \Phi(t),  \\
\label{eqn:h4} h_4(t) &=& b(t) \sin \Phi(t),
\end{eqnarray}
where $a(t)$ and $b(t)$ are the antenna-pattern functions defined by Eqns.~(12) and (13) in Ref. \cite{Jaranowski1998}, and $\Phi(t)$ is the signal phase given by Eqn.~(\ref{eqn:phase}). The four-component model is generally applied in directed continuous-wave searches \cite{Riles2017}. 

\subsection{Nonperpendicular biaxial rotor}
\label{sec:biaxial-case}

We now consider a nonperpendicular biaxial rotor, when $\theta \neq  \pi/2$. The gravitational-wave emission is at both $\fstar$ and $2\fstar$, and the waveform is given by \cite{Jaranowski1998}
\begin{eqnarray}
\label{eqn:h2p} h_{2+} &=& \frac{1}{2}h_0(1+\cos^2\iota)\sin^2\theta\cos 2\Phi ,\\
h_{2\times} &=& h_0 \cos \iota \sin^2\theta \sin 2 \Phi\, \\
h_{1+} &=&\frac{1}{8}h_0\sin 2\iota \sin 2\theta \sin \Phi,\\
\label{eqn:h1c} h_{1\times} &=& \frac{1}{4} h_0 \sin \iota \sin 2 \theta \cos \Phi.
\end{eqnarray}
The four components in Eqns.~(\ref{eqn:h1})--(\ref{eqn:h4}) become eight
\begin{equation}
\label{eqn:sig_form}
h(t)=\sum_{l=1}^{2}\sum_{m=1}^{4}\mathcal{A}_{lm} h_{lm}(t).
\end{equation}
The amplitudes $\mathcal{A}_{lm}$, depending on $h_0$, $\Phi_0$, $\iota$, $\pol$, and $\theta$, are associated with the eight linearly independent components at both $\fstar$ and $2\fstar$
\begin{eqnarray}
h_{l1}(t) &=& a(t) \cos l\Phi(t), \\
h_{l2}(t) &=& b(t) \cos l\Phi(t), \\
h_{l3}(t) &=& a(t) \sin l\Phi(t),  \\
 h_{l4}(t) &=& b(t) \sin l\Phi(t).
 \end{eqnarray}

\subsection{General triaxial nonaligned model}
\label{sec:triaxial-non-aligned}
A general gravitational-wave signal model for a triaxial star ($I_1\neq I_2 \neq I_3$), whose spin axis is not aligned with any principal axis, consists of one additional dimensionless amplitude in addition to Eqn.~(\ref{eqn:h0}) 
\begin{equation}
h_0' = \frac{16 \pi^2 \fstar^2(I_2-I_1)}{r}.
\end{equation}
The components of the gravitational-wave signal are in a more complicated form \cite{Jones2015}
\begin{eqnarray}
	\nonumber h_{2+} &=& \frac{1}{2}(1+\cos^2\iota)\{[h_0'(\sin^2\psi - \cos^2\psi \cos^2\theta)\\
	\label{eqn:general_h2}
	&&-h_0 \sin^2\theta]\cos 2\Phi + h_0' \sin 2 \psi \cos \theta \sin 2\Phi\},\\
	\nonumber h_{2\times} &=& - \cos \iota\{ h_0' \sin 2\psi \cos \theta \cos 2 \Phi \\
	\nonumber && - [ h_0' (\sin^2 \psi - \cos^2 \psi \cos^2 \theta) \\
	&& -h_0 \sin^2\theta] \sin 2 \Phi\}, \\
	\nonumber h_{1+} &=& \frac{1}{4}\sin \iota \cos \iota [h_0' \sin 2\psi \sin \theta \cos \Phi\\
	&& + (h_0' \cos^2\psi - h_0)\sin 2\theta \sin \Phi],\\
	\nonumber h_{1\times} &=& -  \frac{1}{4} \sin \iota [ (h_0' \cos^2 \psi - h_0) \sin 2 \theta \cos \Phi \\
	\label{eqn:general_h1}
	&&- h_0' \sin 2\psi \sin \theta \sin \Phi],
\end{eqnarray}
where $\psi$ is the other orientation angle of the triaxial rotor in the frame of the principal axes in addition to $\theta$.

This triaxial nonaligned model can be regarded as a superposition of two signals from two nonperpendicular biaxial rotors (Sec.~\ref{sec:biaxial-case}). Ref.~\cite{Pitkin2015} demonstrates that it is difficult to distinguish between two signals described by Eqns.~(\ref{eqn:h2p})--(\ref{eqn:h1c}) and Eqns.~(\ref{eqn:general_h2})--(\ref{eqn:general_h1}), even with high signal-to-noise ratio (SNR). We note that Eqns.~(\ref{eqn:general_h2})--(\ref{eqn:general_h1}) are only valid for the pinned superfluid model proposed by Ref.~\cite{Jones2010} without free precession. More generally, the $h_+$ and $h_\times$ signal components from a star involving free precession can be written in the form \cite{Zimmermann1980}
\begin{eqnarray}
\nonumber
h_+ &=& -\frac{1}{r} [(\mathcal{R}_{y\mu} \cos\iota - \mathcal{R}_{z\mu}\sin \iota) \\
\label{eqn:hp} &\times&(\mathcal{R}_{y\nu} \cos\iota - \mathcal{R}_{z\nu}\sin \iota) -\mathcal{R}_{x\mu}\mathcal{R}_{x\nu}]\mathcal{A}_{\mu\nu},\\
\label{eqn:hc} h_\times &=& \frac{2}{r} (\mathcal{R}_{y\mu} \cos\iota - \mathcal{R}_{z\mu}\sin \iota)\mathcal{R}_{x\nu}\mathcal{A}_{\mu\nu},
\end{eqnarray}
where amplitudes $\mathcal{A}_{\mu\nu}$ are functions of angular velocity components in the star's body frame and the three principal moments of inertia, defined by Eqn.~(21) in Ref.~\cite{Zimmermann1980}. 
The amplitudes $\mathcal{A}_{\mu\nu}$ are associated with the rotation matrix $\mathcal{R}$ in terms of the Euler angles $\theta$, $\psi$, and $\varphi$, given by \cite{Zimmermann1980}
\begin{widetext}
\begin{equation}
\mathcal{R} = 
\begin{pmatrix}
\cos \psi \cos\varphi -\cos \theta \sin \psi \sin \varphi& -\sin \psi \cos \varphi - \cos \theta \cos \psi \sin \varphi & \sin \theta \sin \varphi\\
\cos \psi \sin \varphi + \cos \theta \sin \psi \cos \varphi   & -\sin \psi \sin \varphi + \cos \theta \cos \psi \cos \varphi & -\sin \theta \cos \varphi  \\
\sin \theta \sin \psi   & \sin \theta \cos \psi & \cos \theta
\end{pmatrix}.
\end{equation}
\end{widetext}
By substituting $\mathcal{R}$ into Eqns.~(\ref{eqn:hp}) and (\ref{eqn:hc}), the resulting gravitational-wave emission spectrum contains peaks in addition to $\fstar$ and $2\fstar$ \cite{Zimmermann1980,Jones2002,VanDenBroeck2005,Lasky2013}. In the case of small $\theta$, small oblateness, and weak nonaxisymmetry, the first-order contribution peaks are at $2\fstar$ and $\fstar+\fprec \approx \fstar$, where $\fprec$ is the star's precessing frequency \cite{Zimmermann1980}. The second-order contribution peaks appear to be sidelobes of the first-order peaks, e.g., at $2\fstar+2\fprec$ \cite{VanDenBroeck2005}.

In this paper, we focus on comparing a nonperpendicular biaxial rotor (Sec.~\ref{sec:biaxial-case}) to a perpendicular biaxial rotor or a triaxial aligned rotor (Sec.~\ref{sec:triaxial-aligned}; the conventional model adopted in continuous-wave searches). We parameterize the signal waveforms using Eqns.~(\ref{eqn:h2p_classic})--(\ref{eqn:h2c_classic}), and (\ref{eqn:h2p})--(\ref{eqn:h1c}), as described in Ref.~\cite{Jaranowski1998}.\footnote{A reformulation of the waveform parameters is given by Ref.~\cite{Jones2015}, which is adopted in some of the targeted known pulsar searches \cite{Pitkin2015,O2-target}. The two sets of parameters can be transformed interchangeably for comparison purposes.}

\section{Coherent matched filter: $\mathcal{F}$-statistic}
\label{sec:F-stat}
The time-domain data collected by a detector takes the form
\begin{equation}
x(t)= h(t) + n(t),
\end{equation}
where $n(t)$ stands for stationary, additive noise. 
We define a scalar product $(\cdot|\cdot)$ as a sum over single-detector inner products,
\begin{eqnarray}
(x|y) &=& \mathop{\sum} \limits_{X} (x^X|y^X) \\
&=& \mathop{\sum} \limits_{X} 4\Re \int_{0}^{\infty}df \frac{\tilde{x}^X(f)\tilde{y}^{X*}(f)}{S_h^X(f)},
\end{eqnarray}
where $X$ indexes the detector, $S_h^X(f)$ is the single-sided power spectral density (PSD) of detector $X$, the tilde denotes a Fourier transform, and $\Re$ returns the real part of a complex number \cite{Prix2007}. 
The likelihood function of detecting a signal in data $x(t)$ is given by \cite{Jaranowski1998}
\begin{equation}
\label{eqn:total_max_lik} 
\ln \Lambda = (x|h)-\frac{1}{2}(h|h).
\end{equation}
The two frequency components of a gravitational-wave signal given by Eqn.~(\ref{eqn:sig_form}) are in narrow bands around $\fstar$ and $2\fstar$. Hence to a good approximation, we can write \cite{Jaranowski1998}
\begin{equation}
\label{eqn:total_max_lik_approx}
\ln \Lambda \approx (x|h_1)-\frac{1}{2}(h_1|h_1) + (x|h_2)-\frac{1}{2}(h_2|h_2).
\end{equation}
The $\mathcal{F}$-statistic is a frequency-domain estimator maximizing $\ln \Lambda $ with respect to $\mathcal{A}_{lm}$. 

Usually in $\mathcal{F}$-statistic-based searches, it is assumed that the gravitational-wave emission is only at $2\fstar$ (Sec.~\ref{sec:triaxial-aligned}). The $\mathcal{F}$-statistic is expressed in the form
\begin{equation}
\label{eqn:F2}
\mathcal{F}_2 = \frac{1}{2} x_\mu \mathcal{M}^{\mu \nu} x_\nu,
\end{equation}
where we write $x_\mu = (x|{h_2}_\mu)$, and $\mathcal{M}^{\mu \nu}$ denotes the matrix inverse of $\mathcal{M}_{\mu \nu}=({h_2}_\mu|{h_2}_\nu)$.  
Assuming the noise $n(t)$ is Gaussian, the random variable $2\mathcal{F}_2$ follows a central chi-squared distribution with four degrees of freedom without a signal, whose probability density function (PDF) is
\begin{equation}
 p(2\mathcal{F}_2)=\chi^2(2\mathcal{F}_2; 4,0).
\end{equation}
With a signal present in Gaussian noise, the chi-squared distribution of $2\mathcal{F}_2$ is noncentral, viz.
\begin{equation}
\label{eqn:chi2-dist}
p(2\mathcal{F}_2)=\chi^2(2\mathcal{F}_2; 4,\rho_2^2), 
\end{equation}
with noncentrality parameter \cite{Jaranowski1998}
\begin{equation}
\rho_2^2 = \frac{K_2 h_0^2 \Tcoh}{S_h(2\fstar)},
\end{equation}
where the constant $K_2$ depends on $\iota$, the sky location of the source, and the number of detectors, and $\Tcoh$ is the coherent time interval over which $\mathcal{F}_2$ is computed. Here we assume the same single-sided PSD, $S_h(f)$, in all detectors. The optimal SNR equals $\rho_2$. 

To consider a dual-harmonic signal at both $\fstar$ and $2\fstar$ with eight nonindependent amplitudes $\mathcal{A}_{lm}$, the optimal matched filter maximizing Eqn.~(\ref{eqn:total_max_lik_approx}) needs to be obtained through prohibitively expensive numerical calculation, taking into consideration the five parameters ($h_0$, $\Phi_0$, $\iota$, $\pol$, and $\theta$) that $\mathcal{A}_{lm}$ depend on. To make a search computationally feasible, a reduced likelihood function is used to compute the $\mathcal{F}$-statistic, assuming that $\mathcal{A}_{lm}$ are independent with respect to $h_0$, $\Phi_0$, $\iota$, $\pol$, and $\theta$. The two terms in Eqn.~(\ref{eqn:total_max_lik_approx}), $(x|h_1)-\frac{1}{2}(h_1|h_1)$ and $(x|h_2)-\frac{1}{2}(h_2|h_2)$, are maximized independently with respect to $\mathcal{A}_{lm}$ in two separate narrow bands, giving the total $\mathcal{F}$-statistic \cite{Jaranowski1998}
\begin{equation}
\mathcal{F} = \mathcal{F}_1 + \mathcal{F}_2,
\end{equation}
where $\mathcal{F}_1$ is computed in the same way as (\ref{eqn:F2}) but by replacing $h_2$ component with $h_1$. With a dual-harmonic signal present in Gaussian noise and assuming the same $S_h(f)$ in all detectors, the random variable $2\mathcal{F}$ follows a noncentral chi-squared distribution with eight degrees of freedom, and the noncentrality parameter is given by \cite{Jaranowski1998}
\begin{equation}
\rho_0^2 = \rho_1^2 + \rho_2^2,
\end{equation}
where
\begin{equation}
\label{eqn:rho1}
\rho_1^2 = \frac{K_1 h_0^2\Tcoh\sin^2 2\theta}{S_h(\fstar)},
\end{equation}
and
\begin{equation}
\label{eqn:rho2}
\rho_2^2 = \frac{K_2 h_0^2\Tcoh\sin^4 \theta}{S_h(2\fstar)}.
\end{equation}
In Eqns.~(\ref{eqn:rho1}) and (\ref{eqn:rho2}), $K_1$ and $K_2$ both depend on $\iota$, the sky location of the source, and the number of detectors.

In this paper, we leverage the existing, fully tested $\mathcal{F}$-statistic software infrastructure in the LSC Algorithm Library Applications (LALApps)\footnote{https://lscsoft.docs.ligo.org/lalsuite/lalapps/index.html} to compute $\mathcal{F}$ as a function of frequency over $\Tcoh$ \cite{F-stat2011}. The software operates on the raw data collected by the interferometers in the form of short Fourier transforms (SFTs), usually with length $T_{\rm SFT}=30$\,min for each SFT.


\section{Dual-harmonic continuous-wave signal tracking} 
\label{sec:hmm}

\subsection{HMM formulation}
\label{sec:hmm_formulation}
A HMM is a memoryless automaton composed of a hidden (unobservable) state variable $q(t) \in \{q_1, \cdots, q_{N_Q}\}$ and a measurement (observable) variable $o(t)\in \{o_1, \cdots, o_{N_O}\}$ sampled at time $t \in \{t_0, \cdots, t_{N_T}\}$. We use $N_Q$, $N_O$, and $N_T$ to denote the total number of hidden states, observable states, and discrete time steps, respectively. The most probable sequence of hidden states given the observations over total observing time $\Tobs$ is computed by the classic Viterbi algorithm \cite{Viterbi1967}. A full description can be found in Refs. \cite{Suvorova2016} and \cite{Sun2018}.

In a HMM, the emission probability at discrete time $t_n$ is defined as the likelihood of hidden state $q_i$ being observed in state $o_j$, given by \cite{Suvorova2016}
\begin{equation}
\label{eqn:likelihood}
L_{o_j q_i} = P [o(t_n)=o_j|q(t_n)=q_i].
\end{equation}
We set the one-dimensional hidden state variable $q(t)=\fstar(t)$. The discrete hidden states are mapped one-to-one to the frequency bins in the output of a frequency-domain estimator computed over coherent time interval $\Tcoh$. We choose $\Tcoh$ to satisfy
\begin{equation}
\label{eqn:int_T}
\left|\int_t^{t+\Tcoh}dt' \dot{\fstar}(t')\right| < \Delta \fstar
\end{equation}
for $0<t<\Tcoh$, where $\Delta \fstar$ is the frequency bin size in the estimator. At twice the spin frequency of the star, we have
\begin{equation}
\label{eqn:int_T_2f}
\left|\int_t^{t+\Tcoh}dt' 2\dot{\fstar}(t')\right| < 2 \Delta \fstar.
\end{equation}
Here we leverage the existing frequency domain estimator $\mathcal{F}$-statistic described in Sec.~\ref{sec:F-stat}, and define log emission probability computed over each interval [$t,t+\Tcoh$], given by \cite{Jaranowski1998,F-stat2011,Suvorova2016}
\begin{eqnarray}
\label{eqn:emi_prob_matrix}
\ln L_{o(t) q_i} &=& \ln P [o(t)|{\fstar}_i \leq \fstar(t) \leq {\fstar}_i+\Delta \fstar]\\ 
\label{eqn:matrix_propto}
&= & \mathcal{F}_1({\fstar}_i) + \mathcal{F}_2(2{\fstar}_i),
\end{eqnarray}
where ${\fstar}_i$ is the frequency value in the $i$-th bin. We use $\Delta \fstar = 1/(4 \Tcoh)$ and $2\Delta \fstar = 1/(2 \Tcoh)$ as frequency bin sizes when computing $\mathcal{F}_1$ and $\mathcal{F}_2$, respectively, such that both the $\fstar$ and $2\fstar$ signal components stay in one bin for each time interval $\Tcoh$.

The transition probability of the hidden state from discrete time $t_n$ to $t_{n+1}$ is defined as \cite{Suvorova2016}
\begin{equation}
\label{eqn:prob_matrix}
A_{q_j q_i} = P [q(t_{n+1})=q_j|q(t_n)=q_i].
\end{equation}
The choice of $A_{q_j q_i}$ depends on the frequency evolution characteristics of the source. If we consider a scenario that $\fstar$ walks randomly due to timing noise, which is dominant compared to the star's secular spin down or spin up, $A_{q_j q_i}$ takes the form \cite{Suvorova2016}
\begin{equation}
\label{eqn:trans_matrix_sw}
A_{q_{i+1} q_i} = A_{q_i q_i} = A_{q_{i-1} q_i} = \frac{1}{3},
\end{equation}
with all other entries being zero. Or if the timescale of timing noise is much longer than the star's secular spin-down timescale, $A_{q_j q_i}$ is given by \cite{Sun2018}
\begin{equation}
\label{eqn:trans_matrix_sd}
A_{q_{i-1} q_i} = A_{q_i q_i} = \frac{1}{2},
\end{equation}
with all other entries vanishing. 

We choose a uniform prior,
\begin{equation}
\Pi_{q_i} = P [q(t_0)=q_i] = N_Q^{-1}. 
\end{equation}
The probability that the hidden state path $Q=\{q(t_0), \cdots, q(t_{N_T})\}$ gives rise to the observed sequence $O=\{o(t_0), \cdots, o(t_{N_T})\}$ via a Markov chain equals
\begin{equation}
\label{eqn:prob}
\begin{split}
P(Q|O) = & L_{o(t_{N_T})q(t_{N_T})} A_{q(t_{N_T})q(t_{N_T-1})} \cdots L_{o(t_1)q(t_1)} \\ 
& \times A_{q(t_1)q(t_0)} \Pi_{q(t_0)}.
\end{split}
\end{equation}
The most probable path maximizes $P(Q|O)$, denoted by
\begin{equation}
Q^*(O)= \arg\max P(Q|O),
\end{equation}
where $\arg \max (\cdots)$ returns the argument that maximizes the function $(\cdots)$. $Q^*(O)$ gives the best estimate of $q(t)$ over the total observation $\Tobs = N_T \Tcoh$.

\subsection{Path probability distribution versus SNR}
\label{sec:analytic}

We now compare the distributions of path probabilities between tracking $2\fstar$ only and tracking $\fstar$ and $2\fstar$ simultaneously, when a dual-harmonic signal is present. For simplicity, we assume stationary, Gaussian noise, and hence $\mathcal{F}$-statistic is independently and identically distributed. For $\mathcal{F}=\mathcal{F}_2$, the random variable $2\mathcal{F}$ computed over each block of $\Tcoh$ is chi-squared distributed with four degrees of freedom. If $Q^*(O)$ does not intersect the true signal path anywhere, the PDF of $z=\ln P(Q|O)$ is given by \cite{Jaranowski1998,Sun2018}
\begin{equation}
\label{eqn:N-chi2-dist-noise}
p(z) = \chi^2\left(z; 4 N_T,0\right).
\end{equation}
If $Q^*(O)$ coincides exactly with the true signal path, we have \cite{Jaranowski1998,Sun2018}
\begin{equation}
	\label{eqn:N-chi2-dist}
	p(z) = \chi^2\left[z; 4 N_T,\frac{K_2 h_0^2\Tobs \sin^4 \theta}{S_h(2\fstar)}\right].
\end{equation}
If both $\fstar$ and $2\fstar$ components are tracked, the variable $2\mathcal{F} = 2\mathcal{F}_1 +2\mathcal{F}_2$ computed over each block of $\Tcoh$ is chi-squared distributed with eight degrees of freedom. The PDFs in Eqns.~(\ref{eqn:N-chi2-dist-noise}) and (\ref{eqn:N-chi2-dist}) become \cite{Jaranowski1998,Sun2018}
\begin{equation}
\label{eqn:N-chi2-dist-noise-f2f}
p(z) = \chi^2\left(z; 8 N_T,0\right),
\end{equation}
and
\begin{equation}
\label{eqn:N-chi2-dist-f2f}
p(z) = \chi^2\left[z; 8 N_T,\frac{K_1 h_0^2\Tobs\sin^2 2\theta}{S_h(\fstar)} + \frac{K_2 h_0^2\Tobs\sin^4 \theta}{S_h(2\fstar)}\right].
\end{equation}

\begin{figure}[!tbh]
	\centering
	\includegraphics[width=\columnwidth]{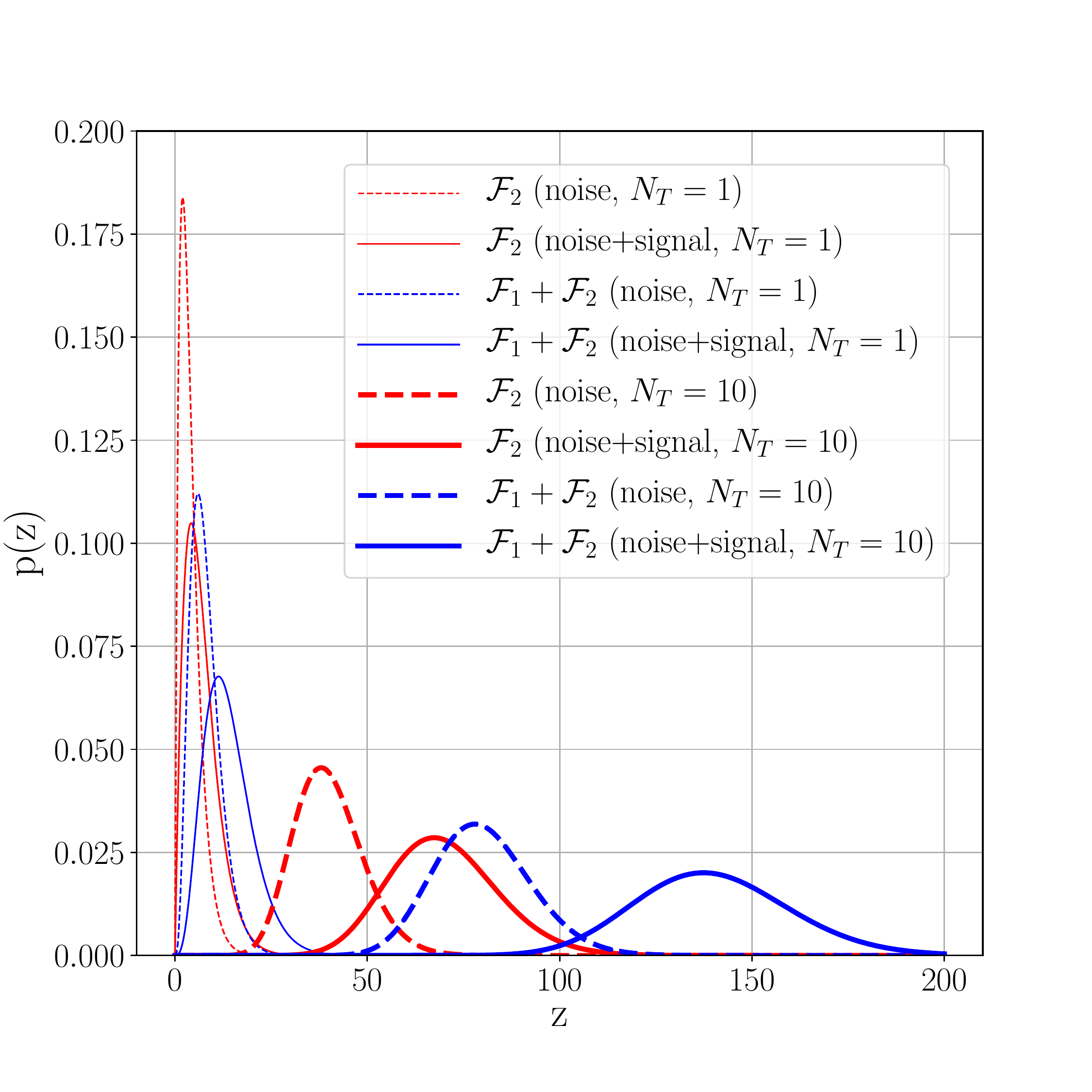}
	\caption[]{Probability density function $p(z)$ for log likelihood $z = \ln P(Q|O)$ along path $Q$. The red and blue curves indicate tracking $2\fstar$ only and tracking both $\fstar$ and $2\fstar$, respectively. The thin and thick curves indicate the number of tracking steps $N_T=1$ and $N_T=10$, respectively. The solid and dashed curves indicate that $Q$ intersects perfectly and not at all with the true signal path, respectively. The optimal Viterbi path obtained can overlap partly with the true signal path, yielding a distribution in between the solid and dashed curves. 
	A signal is more distinguishable from noise by tracking both $\fstar$ and $2\fstar$ than tracking $2\fstar$ only. As more steps $N_T$ are taken, it is progressively easier to distinguish a signal from noise. The search cost increases approximately $\propto N_T$ for both methods (see Sec.~\ref{sec:discussion}).
	Parameters: $\rho_1^2 =\rho_2^2 = 3$.}
	\label{fig:pdf}
\end{figure}

Figure \ref{fig:pdf} shows distributions of path probabilities for tracking single component ($\mathcal{F} =\mathcal{F}_2$; red curves) and both components ($\mathcal{F} =\mathcal{F}_1+\mathcal{F}_2$; blue curves). The blue dashed and solid curves display $p(z)=\chi^2(z; 8N_T, 0)$ (pure noise path) and $p(z)=\chi^2(z; 8N_T, N_T \rho_0^2)$ (true signal path), respectively. Similarly, the red dashed and solid curves display $p(z)=\chi^2(z; 4N_T, 0)$ and $p(z)=\chi^2(z; 4N_T, N_T \rho_2^2)$, respectively. The thin and thick curves indicate $N_T = 1$ and $N_T = 10$, respectively. In this example, we show an optimal scenario with $\rho_1^2 = \rho_2^2$. The figure demonstrates that it is much easier to distinguish a signal from noise by tracking both components. Increasing $N_T$ can always make the distribution of signal paths more significantly differ from that of noise paths, for both methods. Note that in reality, the number of steps that the optimal Viterbi path intersects the true signal path depends on SNR, which is always between 0 and $N_T$. Hence the distribution of path probabilities in fact lies somewhere between the dashed and solid curves. The true PDF of Viterbi paths is difficult to compute mathematically. An analytic approximation of the true PDF is discussed in Ref.~\cite{Suvorova2017}. The search cost increases approximately $\propto N_T$ for both methods. A detailed discussion about computing cost is provided in Sec.~\ref{sec:discussion}.

\section{Simulation and sensitivity}
\label{sec:simulation}

In this section, we begin with a detailed example, demonstrating the sensitivity improvement obtained from dual-harmonic tracking (Sec.~\ref{sec:example}). We define detection statistics and calculate the threshold in Sec.~\ref{sec:viterbi_score}. In Sec.~\ref{sec:efficiency}, we adopt the threshold for a given false alarm probability, carry out Monte Carlo simulations, and study the rates of correctly recovering injected signals, i.e., detection efficiency, for various $h_0$, $\theta$ and $\cos\iota$ values. 

\subsection{Tracking example}
\label{sec:example}

\begin{table}[!tbh]
	\centering
	\setlength{\tabcolsep}{5pt}
	\renewcommand\arraystretch{1.5}
	\begin{tabular}{lll}
		\hline
		Injection parameters & Symbol & Value \\
		\hline
		Right ascension & $\alpha$ & 23h 23m 26.0s\\
		Declination & $\delta$ & $58^{\circ}48' 0.0''$\\
		Detector PSD &$S_h (f)^{1/2}$ & $4 \times 10^{-24}$\,Hz$^{-1/2}$ \\
		Initial spin frequency & $f_\star$ & 100.1\,Hz \\
		\hline
		Search parameters & Symbol & Value \\
		\hline
		Total observing time & $T_{\rm obs}$ & 50\,d\\
		Coherent time & $\Tcoh$ & 5\,d \\
		Number of steps & $N_T$ & 10\\
		\hline
	\end{tabular}
	\caption[parameters]{Injection parameters used to create the synthetic data and search parameters.}
	\label{tab:inj-paras}
\end{table}

We start by showing one representative example of dual-harmonic tracking. We firstly generate a set of synthetic data for $\Tobs=50$\,d at two detectors (the LIGO Hanford and Livingston observatories) using \textit{Makefakedata} version 4 from LALApps, containing a dual-harmonic signal from a nonperpendicular biaxial rotor (Sec.~\ref{sec:biaxial-case}). The source sky position, detector PSD, and initial $\fstar$ are shown in the top half of Table~\ref{tab:inj-paras}. In this example, we set $h_0 = 8 \times 10^{-26}$, $\theta = 30^\circ$, and $\cos \iota = 0.75$, corresponding to $h_{2+} = 1.56\times10^{-26}$, $h_{2\times} = 1.50\times10^{-26}$, $h_{1+} = 8.59\times10^{-27}$, and $h_{1\times} = 1.15\times10^{-26}$ using Eqns.~(\ref{eqn:h2p})--(\ref{eqn:h1c}), and randomly choose $\psi_{\rm pol} = 0.93$\,rad and $\Phi_0 = 1.19$\,rad. 
Here we assume a scenario where the signal frequency wanders stochastically due to timing noise. We approximate the spin wandering by an unbiased random walk or Wiener process, and let $\fstar$ jump randomly anywhere within $\pm \Delta \fstar = 5.787 \times 10^{-7}$\,Hz with uniform probability every five days (following the strategy described in Ref.\cite{Suvorova2016}). The search is conducted by tracking $N_T = 10$ consecutive coherent intervals, with each lasting for $\Tcoh = 5$\,d (see the bottom half of Table~\ref{tab:inj-paras}), in three ways: (a) tracking $\fstar$ only, (b) tracking $2\fstar$ only, and (c) tracking both $\fstar$ and $2\fstar$ simultaneously.

\begin{figure*}
	\centering
	\subfigure[]
	{
		\label{fig:sample-f}
		\scalebox{0.38}{\includegraphics{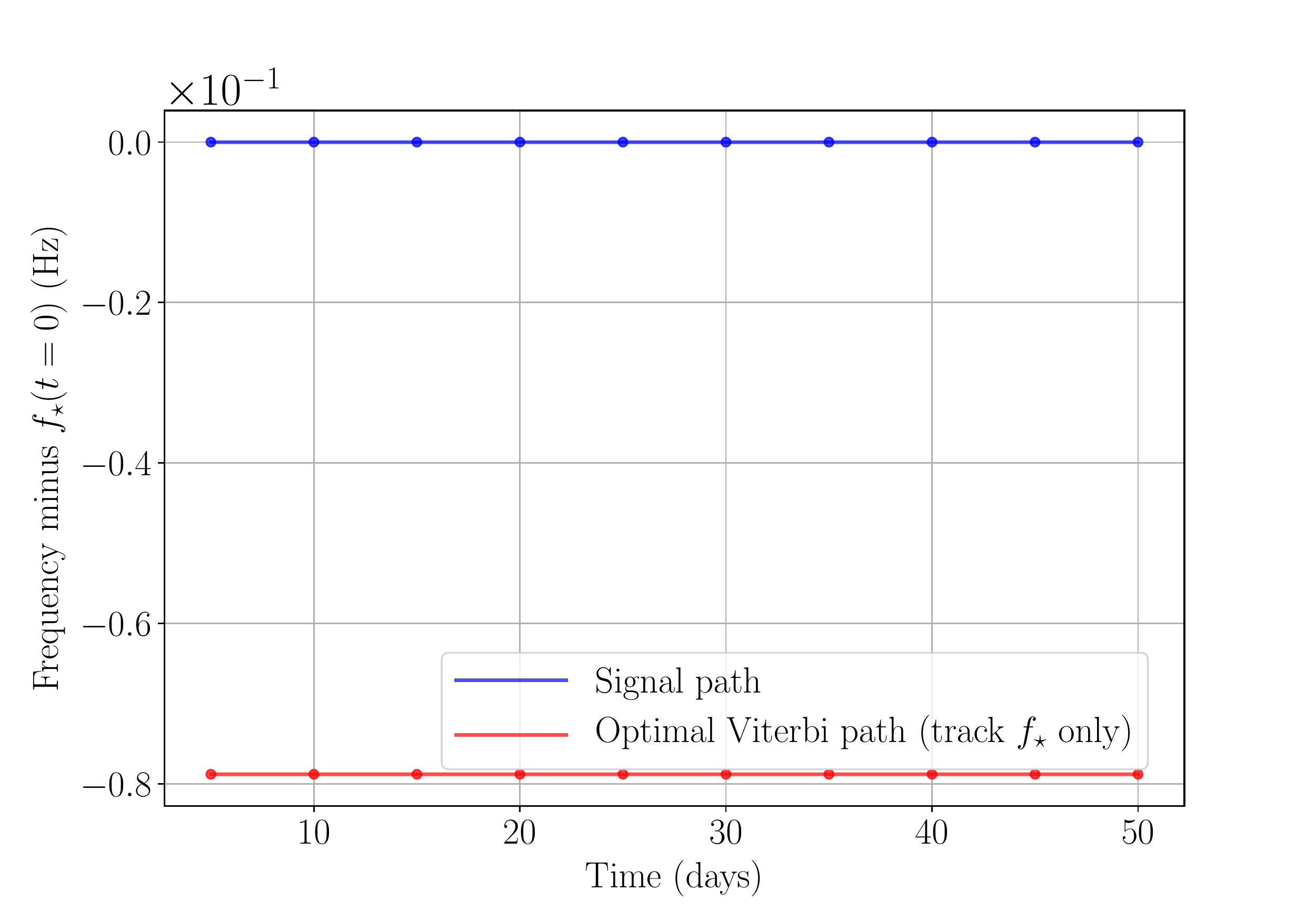}}
	}
	\subfigure[]
	{
		\label{fig:sample-2f}
		\scalebox{0.38}{\includegraphics{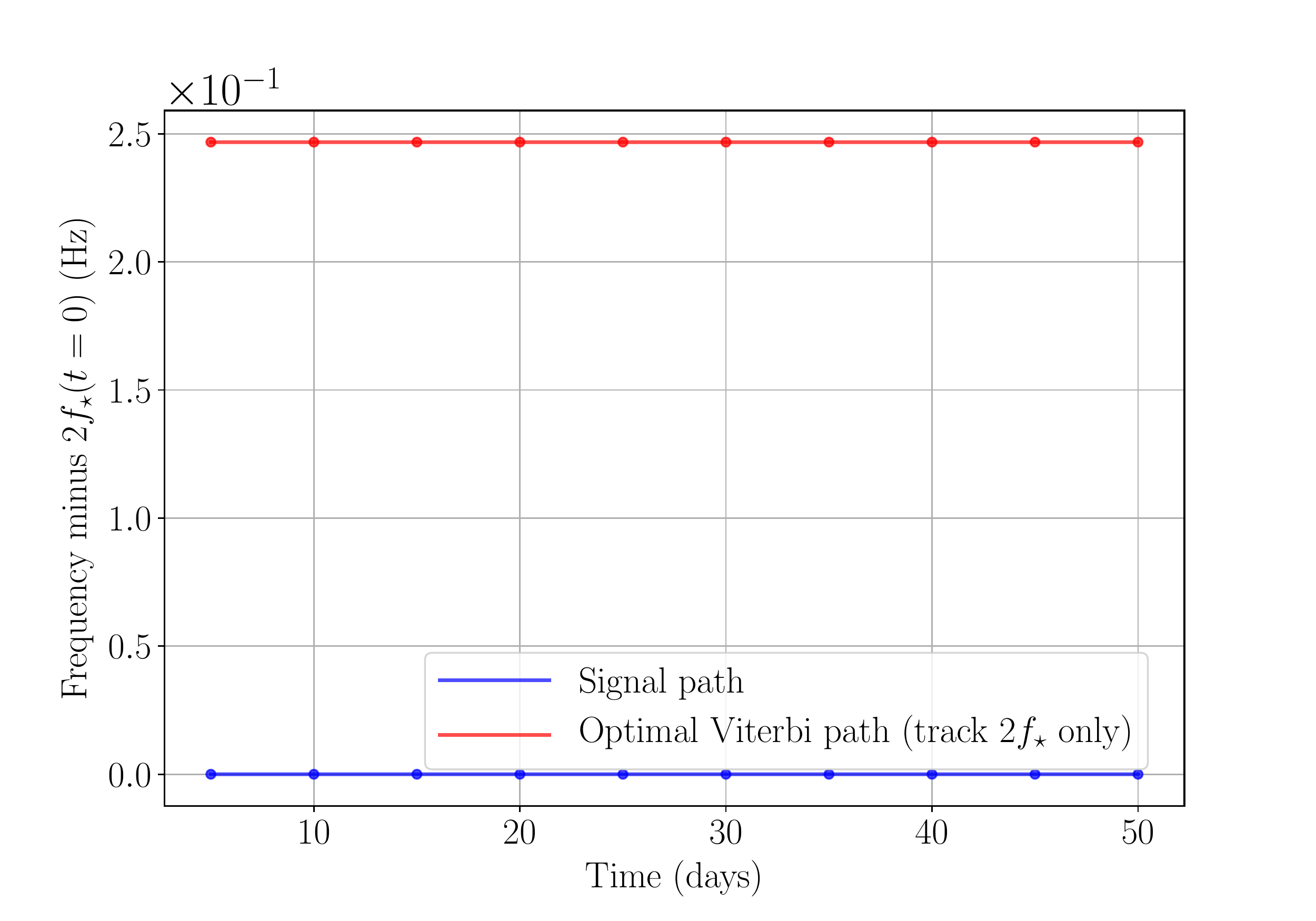}}
	}
	\subfigure[]
	{
		\label{fig:sample-f2f}
		\scalebox{0.38}{\includegraphics{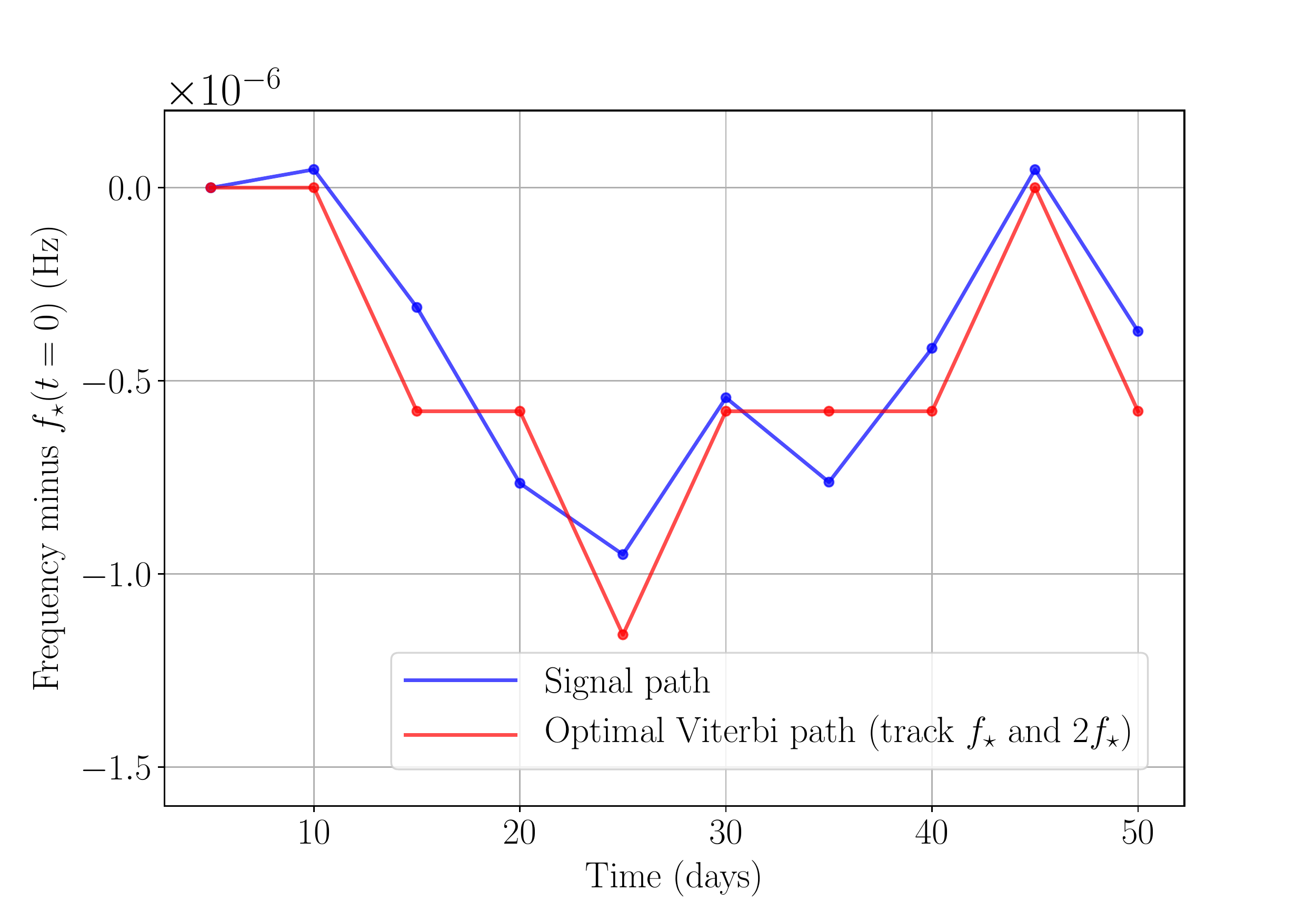}}
	}
	\caption[]{Injected signal paths $\fstar(t)$ (blue curves) and optimal Viterbi paths (red curves). Panels (a)--(c) display the results for tracking $\fstar$ only, tracking $2\fstar$ only, and tracking both $\fstar$ and $2\fstar$, respectively. The injection cannot be recovered in either (a) or (b). A good match is obtained in (c), with RMSE $=1.6 \times 10^{-7}$\,Hz, corresponding to $0.28 \Delta \fstar$. The fluctuation of the signal frequency is too small to be seen in (a) and (b).}
	\label{fig:sample}
\end{figure*}

Figure~\ref{fig:sample} displays the tracking results. The blue and red curves indicate the injected signal paths and optimal Viterbi paths returned from the tracking, respectively. Panels (a)--(c) correspond to the above tracking methods (a)--(c), respectively. It is demonstrated that only by tracking both $\fstar$ and $2\fstar$, the injection can be recovered accurately. The root-mean-square error (RMSE) between the optimal Viterbi path and injected signal path in (c) is $1.6 \times 10^{-7}$\,Hz (i.e., $0.28 \Delta \fstar$). The error is introduced mainly because the HMM takes discrete values of $\fstar$ with $\Delta \fstar$ as the smallest step size, while the injected $\fstar(t)$ can take any value within a bin. Note that the frequency fluctuations are too small to be seen in panels (a) and (b). The three blue curves in (a)--(c) are in the same shape. The red curves in (a) and (b) also fluctuate.

\subsection{Viterbi score and threshold}
\label{sec:viterbi_score}

In order to quantify the improvement in detection efficiency, $1-P_{\rm d}$, where $P_{\rm d}$ is the false dismissal probability, we define the Viterbi score and derive a detection threshold for a given false alarm probability, $P_{\rm a}$. We adopt the definition of Viterbi score in \cite{Sun2018}, given by 
\begin{equation}
\label{eqn:viterbi_score}
S = \frac{\ln \delta_{q^*}{(t_{N_T})} -\mu_{\ln \delta}(t_{N_T})}{\sigma_{\ln \delta}(t_{N_T})}
\end{equation}
with
\begin{equation}
\mu_{\ln \delta}(t_{N_T}) = N_Q^{-1} \sum_{i=1}^{N_Q} \ln \delta_{q_i}(t_{N_T})
\end{equation}
and
\begin{equation}
\sigma_{\ln \delta}(t_{N_T})^2 = N_Q^{-1} \sum_{i=1}^{N_Q} [\ln \delta_{q_i}(t_{N_T}) - \mu_{\ln \delta}(t_{N_T}) ]^2,
\end{equation}
where $\delta_{q_i}(t_{N_T})$ denotes the maximum probability of the path ending in state $q_i$ ($1\leq i \leq N_Q$) at step $N_T$, and $\delta_{q^*}{(t_{N_T})}$ is the likelihood of the optimal Viterbi path, i.e. $P[Q^*(O)|O]$. In other words, Viterbi score $S$ is defined, such that the log likelihood of the optimal Viterbi path equals the mean log likelihood of all paths plus $S$ standard deviations at the final step $N_T$. 

Given a choice of $P_{\rm a}$, the detection is deemed successful if $S$ exceeds a threshold $S_{\rm th}$. The value of $S_{\rm th}$ varies with $N_T$, $N_Q$, the entries in $A_{q_j q_i}$, and weakly depends on the distribution of $L_{o_j q_i}$. Systematic Monte Carlo simulations are always required in practice to calculate $S_{\rm th}$ for each HMM implementation. We normally divide the full frequency band into multiple 1-Hz sub-bands to allow parallelized computing in a real search \cite{ScoX1ViterbiO1,Sun2018}. In this section, we compare the performance of three methods: tracking $\fstar$ only, tracking $2\fstar$ only, and tracking both $\fstar$ and $2\fstar$. Since we use bin sizes $\Delta \fstar$ and $2\Delta \fstar$ for $\fstar$ and $2\fstar$ components, respectively (see Sec.~\ref{sec:hmm_formulation}), we consider a sample 1-Hz sub-band (200--201\,Hz) for $2\fstar$ and a half-Hz sub-band (100--100.5\,Hz) for $\fstar$, such that the total number of hidden states $N_Q$ remains the same for three methods.

We set $P_{\rm a} = 1\%$ and determine $S_{\rm th}$ for each of the three methods by conducting searches on data sets containing pure Gaussian noise. The procedure is as follows. We generate $10^3$ noise realizations for two LIGO detectors with $S_h (f)^{1/2}=4 \times 10^{-24}$\,Hz$^{-1/2}$ for $\Tobs = 50$\,d, set $\Tcoh=5$\,d, adopt $A_{q_j q_i}$ in Eqn.~(\ref{eqn:trans_matrix_sw}) assuming a random walk model, and conduct (a) $\fstar$ only tracking in band 100--100.5\,Hz, (b) $2\fstar$ only tracking in band 200--201\,Hz, and (c) dual-harmonic tracking combining both sub-bands. For each method, the value of $S$ yielding a fraction $P_{\rm a}$ of positive detections is $S_{\rm th}$. We obtain $S_{\rm th} = 7.6663$, 7.8798, and 7.2301 for (a), (b), and (c) respectively. Theoretically speaking, $S_{\rm th}$ values for (a) and (b) should be identical, because we have the same $N_T$, $N_Q$, and $A_{q_j q_i}$, and the noise only $\mathcal{F}$-statistic follows a central chi-squared distribution with four degrees of freedom in both (a) and (b). Empirically, the $\mathcal{F}$-statistic output can be weakly impacted by frequency and noise normalization using different bin sizes \cite{F-stat2011}. Hence we see a small difference between thresholds of (a) and (b), with an error $<3\%$. 

\begin{figure*}[!tbh]
	\centering
	\subfigure[]
	{
		\label{fig:f1-1e-25}
		\scalebox{0.32}{\includegraphics{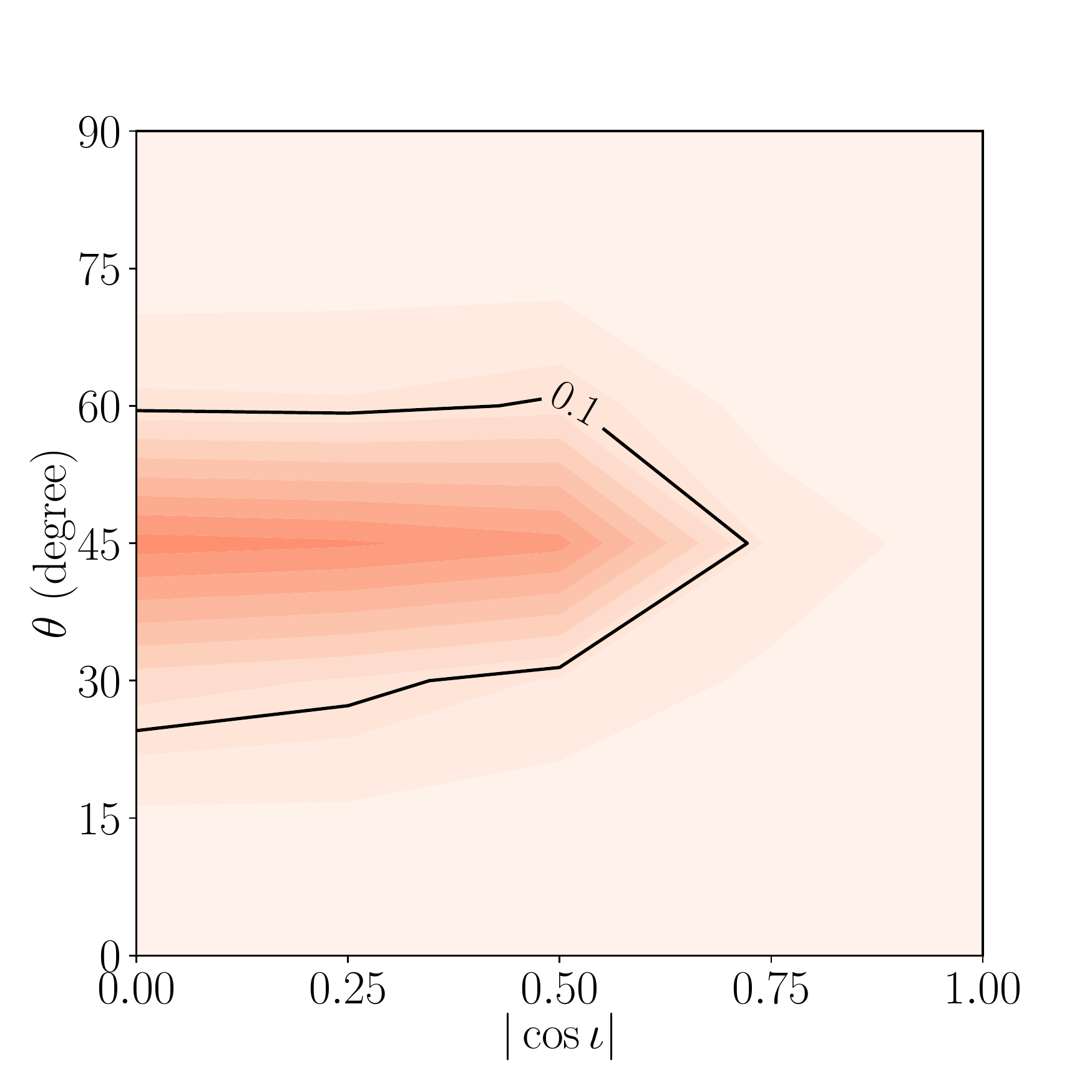}}
	}
	\subfigure[]
	{
		\label{fig:f2-1e-25}
		\scalebox{0.32}{\includegraphics{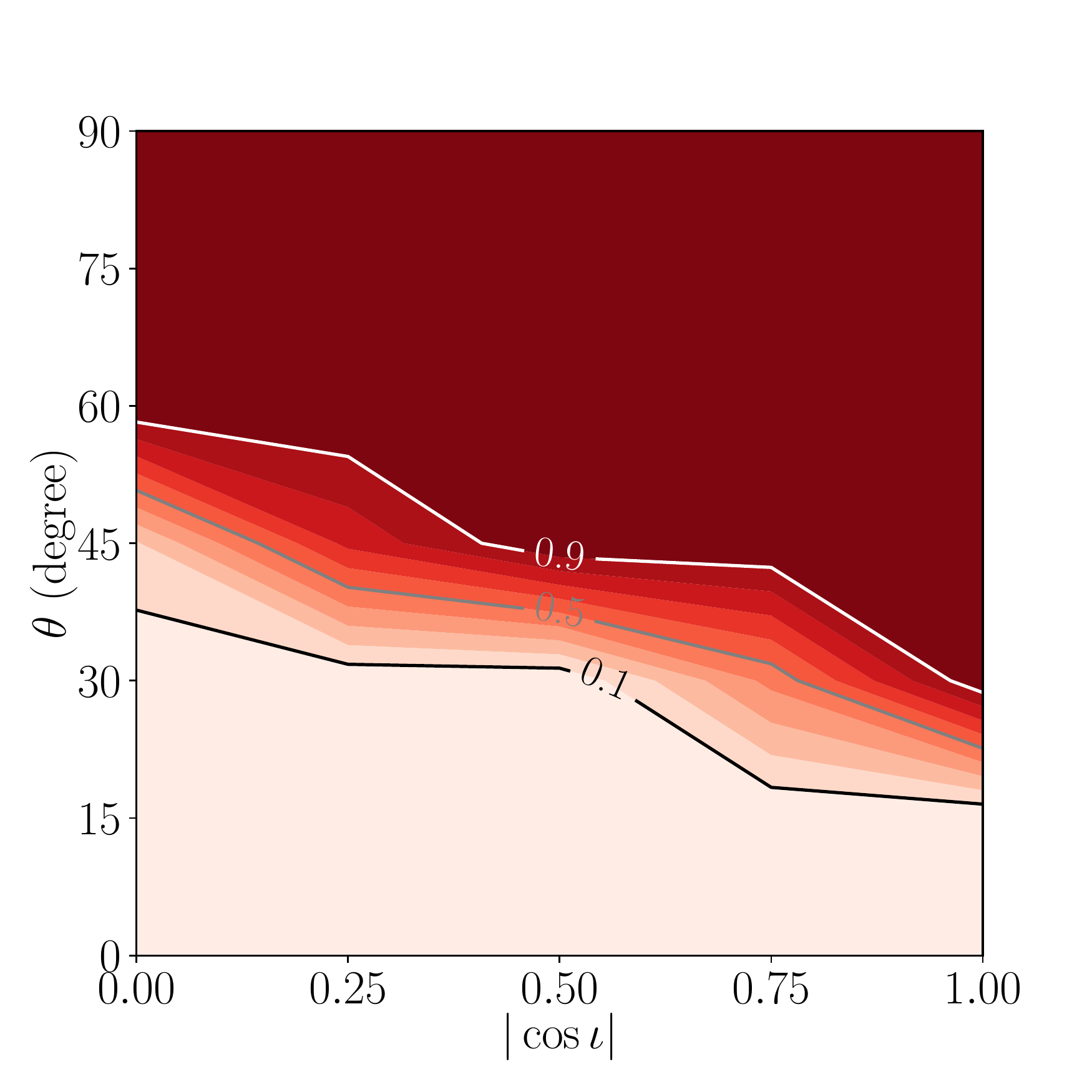}}
	}
	\subfigure[]
	{
		\label{fig:f1+f2-1e-25}
		\scalebox{0.32}{\includegraphics{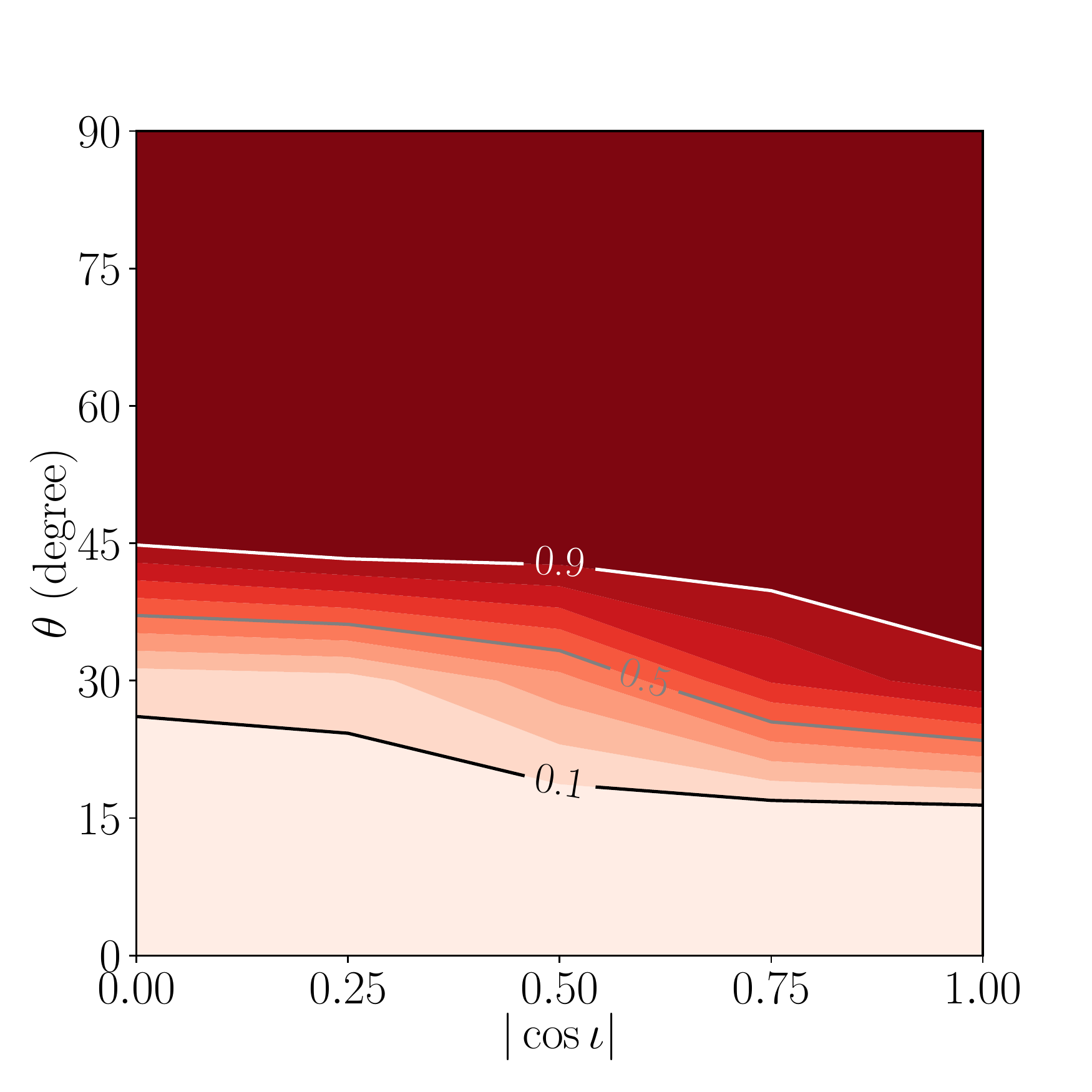}}
	}
	\subfigure[]
	{
	\label{fig:diff_2f-1e-25}
	\scalebox{0.32}{\includegraphics{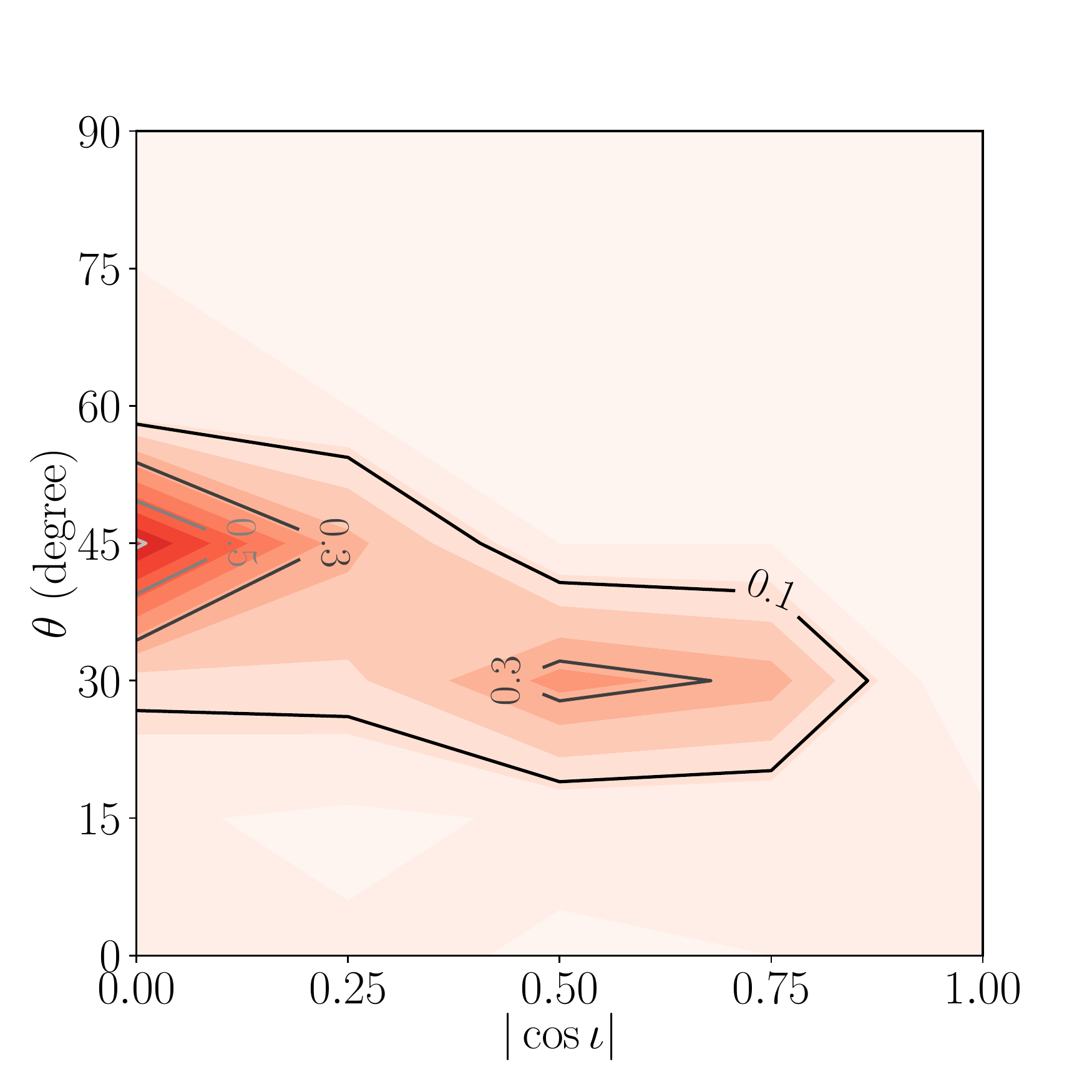}}
    }
	\caption[]{Detection efficiency contours as a function of $|\cos \iota|$ and $\theta$ by tracking (a) $\fstar$ only, (b) $2\fstar$ only, and (c) both $\fstar$ and $2\fstar$ simultaneously. Panel (d) displays the difference between (b) and (c), i.e., the improvement of tracking two frequencies compared to tracking $2\fstar$ only. Parameters: $h_0 = 1\times 10^{-25}$, $S_h (f)^{1/2} = 4 \times 10^{-24}$\,Hz$^{-1/2}$, $\Tcoh = 5$\,d,  $\Tobs = 50$\,d.}
	\label{fig:det_eff_1e-25}
\end{figure*}

\subsection{Detection efficiency}
\label{sec:efficiency}

We now inject synthetic signals in Gaussian noise to study the detection efficiencies of the three tracking methods with $S_{\rm th}$ obtained in Sec.~\ref{sec:viterbi_score}. 
In a real search, since we normally run the tracking in 1-Hz sub-bands, where the interferometric noise PSD can be regarded as flat, the threshold in real interferometric noise does not vary much from Gaussian noise. The sub-bands containing loud instrumental artifacts will be eventually vetoed. A study has been conducted in Ref.\cite{ScoX1ViterbiO1}, comparing the thresholds obtained from Gaussian noise and real O1 data. The resulting $S_{\rm th}$ values match each other with an error $\lesssim 3\%$. The study, however, indicates that the search sensitivity degrades in real interferometric data due to duty cycles and non-Gaussianity, increasing the stain amplitude required for yielding 95\% detection efficiency by a factor of $<2$ \cite{ScoX1ViterbiO1}.
In addition, $S_h (f)$ is a function of frequency in real interferometric data, and hence impacts the SNR in the two frequency bands searched simultaneously using dual-harmonic tracking. Studies of the interferometric $S_h (f)$ and its impact on search sensitivity are needed in a real search.
In this paper, we assume the detector PSD to be identical in the frequency bands tested. We continue using the injection parameters and search configurations in Table~\ref{tab:inj-paras}. 

In the first set of simulations, we set $h_0 = 1\times10^{-25}$, and calculate $h_{2+}$, $h_{2\times}$, $h_{1+}$, and $h_{1\times}$ using Eqns.~(\ref{eqn:h2p})--(\ref{eqn:h1c}) on a grid of $\theta$ and $|\cos\iota|$. For each combination of $\theta \in \{0,15,30,45,60,75,90\}$\,deg and $|\cos \iota| \in \{0,0.25,0.5,0.75,1\}$, we inject 200 signals with both $\psi_{\rm pol}$ and $\Phi_0$ randomly chosen with a uniform distribution within the range $[0,2\pi]$\,rad. The injected $\fstar(t)$ jumps randomly within $\pm \Delta \fstar = 5.787 \times 10^{-7}$\,Hz for every five days. Figure~\ref{fig:det_eff_1e-25} displays the detection efficiency contours of the three methods on the plane of $(\theta, |\cos \iota|)$. Panels (a)--(c) represent results from tracking $\fstar$ only, tracking $2\fstar$ only, and tracking both frequencies, respectively. Darker color stands for higher detection efficiency. The $2\fstar$ component dominates at higher $\theta$ values, and hence the $\fstar$ component contributes little to the sensitivity there. However, at lower $\theta$ values where the $2\fstar$ emission gets weaker, the $\fstar$ component, although generally too weak to be detectable on its own [Figure~\ref{fig:f1-1e-25}], significantly improves the detectability when combined with the $2\fstar$ component. To clearly show the contribution of the weak $\fstar$ component, we plot the improvement from (b) to (c) in panel (d), i.e., the gain by including $\fstar$ component in the tracking. The most significant gain occurs at $20^\circ \lesssim \theta \lesssim 60^\circ$, improving the detection efficiency by $\sim 10\%$ to 70\%. At $\theta\approx 45^\circ$, the detection efficiency can be improved from 19\% to 91\%.

\begin{figure*}[!tbh]
	\centering
	\subfigure[]
	{
		\label{fig:f1-2e-25}
		\scalebox{0.32}{\includegraphics{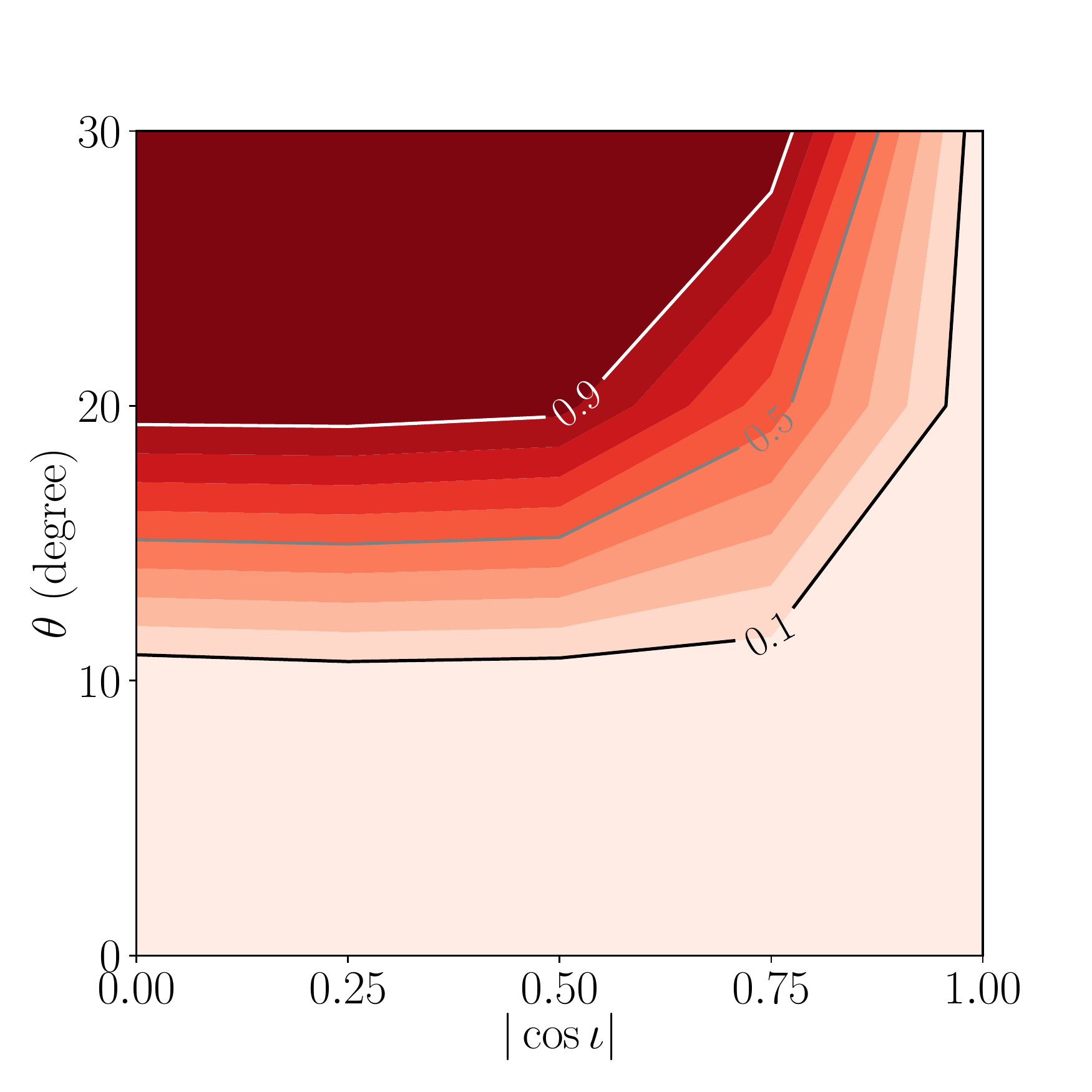}}
	}
	\subfigure[]
	{
		\label{fig:f2-2e-25}
		\scalebox{0.32}{\includegraphics{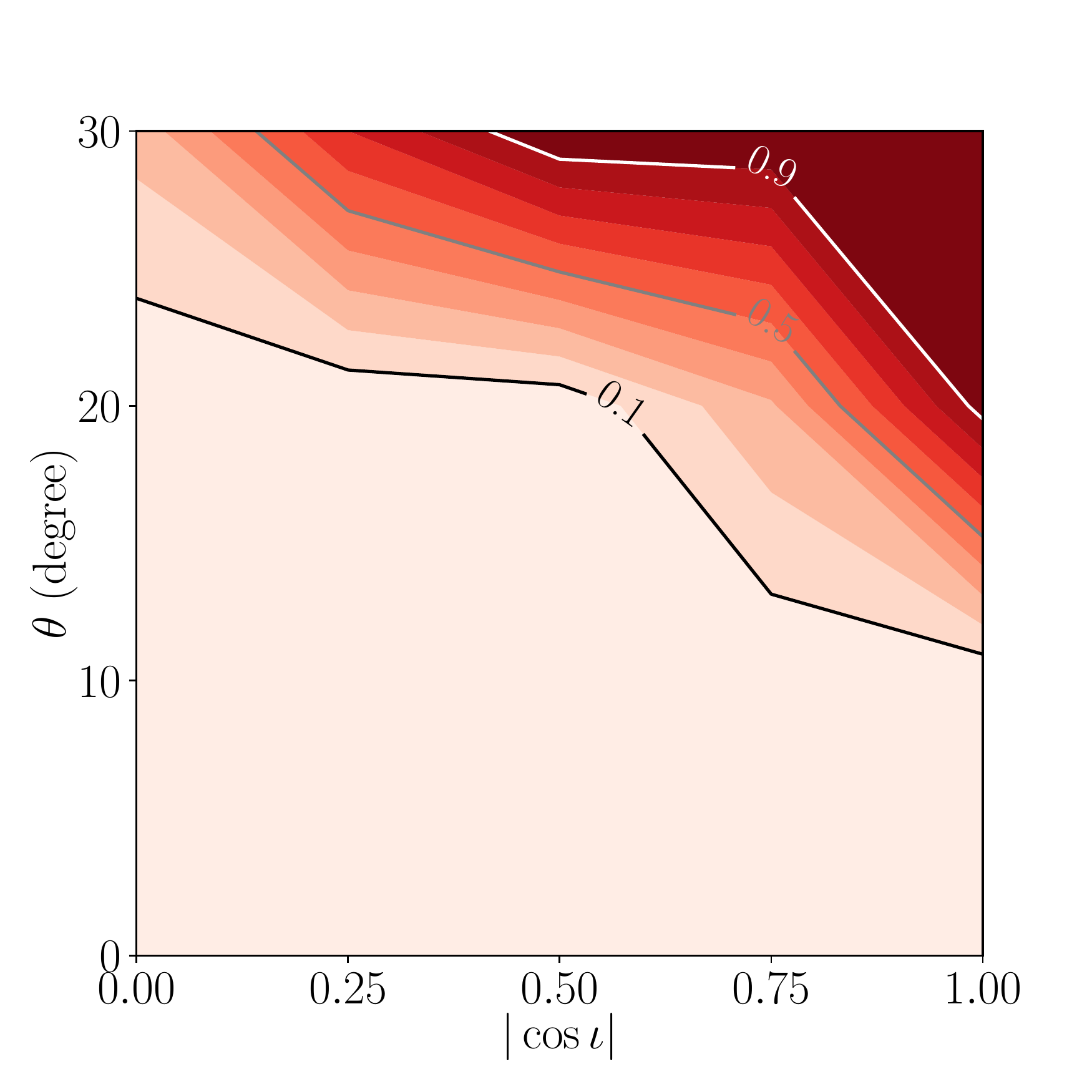}}
	}
	\subfigure[]
	{
		\label{fig:f1+f2-2e-25}
		\scalebox{0.32}{\includegraphics{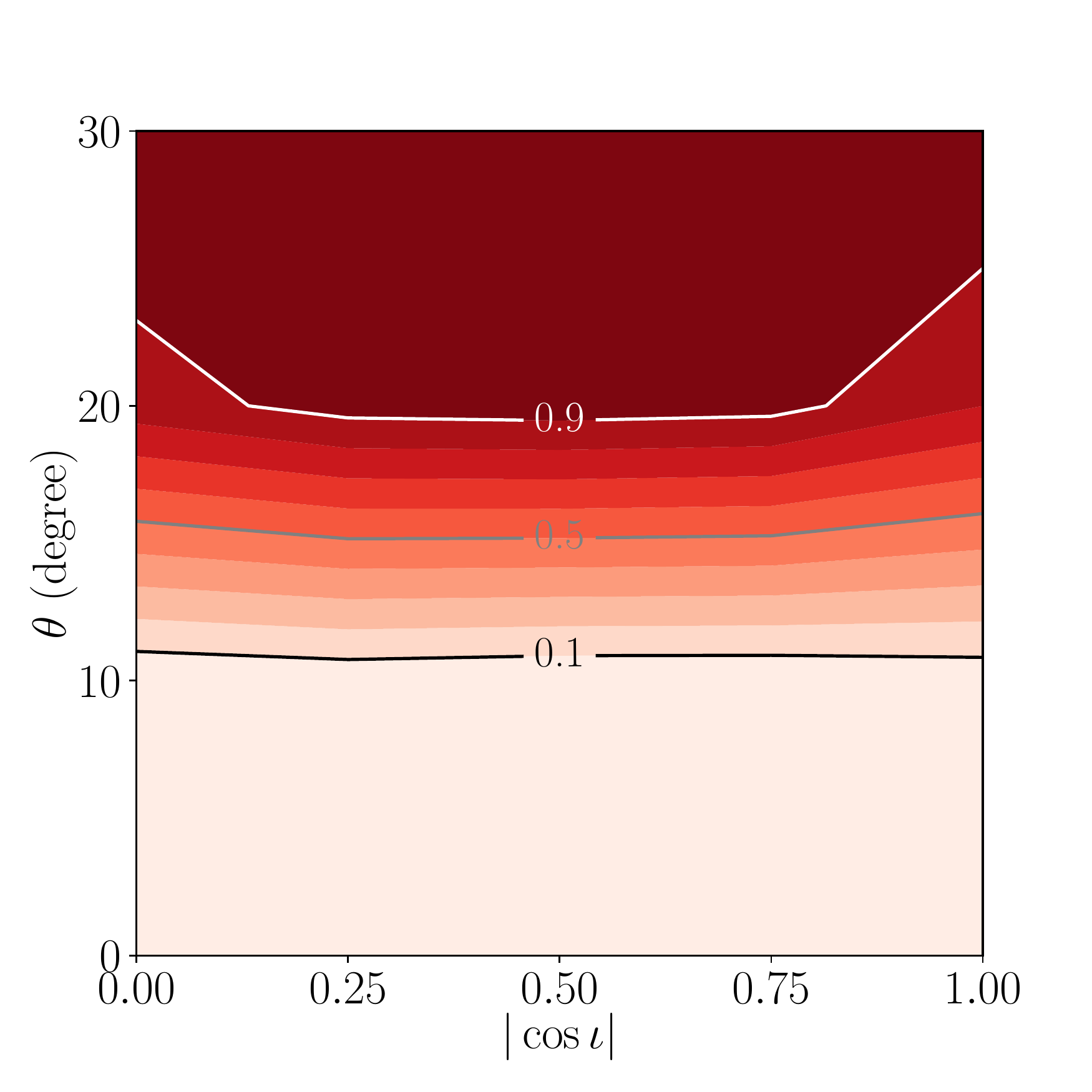}}
	}
	\caption[]{Detection efficiency contours as a function of $|\cos \iota|$ and $\theta$ by tracking (a) $\fstar$ only, (b) $2\fstar$ only, and (c) both $\fstar$ and $2\fstar$ simultaneously. Parameters: $h_0 = 2\times 10^{-25}$, $S_h (f)^{1/2} = 4 \times 10^{-24}$\,Hz$^{-1/2}$, $\Tcoh = 5$\,d,  $\Tobs = 50$\,d.}
	\label{fig:det_eff_2e-25}
\end{figure*}

In the second set of simulations, we probe the parameter space where the $\fstar$ component dominates, i.e., lower $\theta$ values. For each combination of $\theta \in \{0,10,20,30\}$\,deg and $|\cos \iota| \in \{0,0.25,0.5,0.75,1\}$, we run 200 injections with $h_0 = 2\times10^{-25}$. The other parameters and configurations are the same as the first set. The results are shown in Figure~\ref{fig:det_eff_2e-25}. When $\theta \rightarrow 0^\circ$, the strain amplitudes of $\fstar$ and $2\fstar$ components, scaling as $\theta$ and $\theta^2$, respectively, are both too small to be detectable. For $10^\circ \lesssim \theta \lesssim 30^\circ$, the $\fstar$ only tracking and $2\fstar$ only tracking perform well ($\gtrsim 90\%$ detection efficiency) at lower and higher $|\cos \iota|$ values, respectively, while the dual-harmonic tracking can generally produce detection efficiency better than or similar to any of the single frequency tracking methods. The best improvement from dual-harmonic tracking is achieved for $|\cos \iota| \sim 0.75$, increasing the detection efficiency by $\sim 10\%$--30\% compared to either of the single frequency tracking methods.

The above simulations demonstrate that the dual-harmonic tracking performs significantly better in the parameter space where the strain amplitudes of $\fstar$ and $2\fstar$ are comparable, e.g., at the same order of magnitude. 
In other parameter space where one component is dominant, either $\fstar$ or $2\fstar$, the dual-harmonic tracking still performs generally as good as the single frequency tracking. 
However, when one frequency component vanishes, e.g., $|\cos \iota| \to 1$ or $\theta \to \pi/2$, and the other is at low SNR, we find that dual-harmonic tracking performs slightly worse than tracking a single frequency component, losing $\sim 10\%$ detection efficiency at most (e.g., see the parameter space $|\cos \iota|\approx1$ in Figures.~\ref{fig:det_eff_1e-25} and \ref{fig:det_eff_2e-25}).
This is because by tracking two frequency bands simultaneously at low SNR, while the signal only exists in one band, pure noise is introduced from the band corresponding to the vanishing component. 
In this case, the conventional single frequency tracking remains a better method.
The combination of single frequency tracking and dual-harmonic tracking is necessary in order to obtain the optimal sensitivity in the whole parameter space. 
Here we study the optimal choice of tracking methods as a function of $|\cos \iota|$ and $\theta$. Note that in a real directed search without prior knowledge of $\fstar$, we do not differentiate $2\fstar$ tracking and $\fstar$ tracking. In other words, they are both covered in the conventional single component tracking over the full frequency band. Hence we only compare between the conventional single component tracking and dual-harmonic tracking.
Without knowing the intrinsic parameters of the source, $|\cos \iota|$ and $\theta$, we discuss the cost of conducting two searches using both methods in Sec.~\ref{sec:discussion}.

\begin{figure}[!tbh]
	\centering
	\includegraphics[width=\columnwidth]{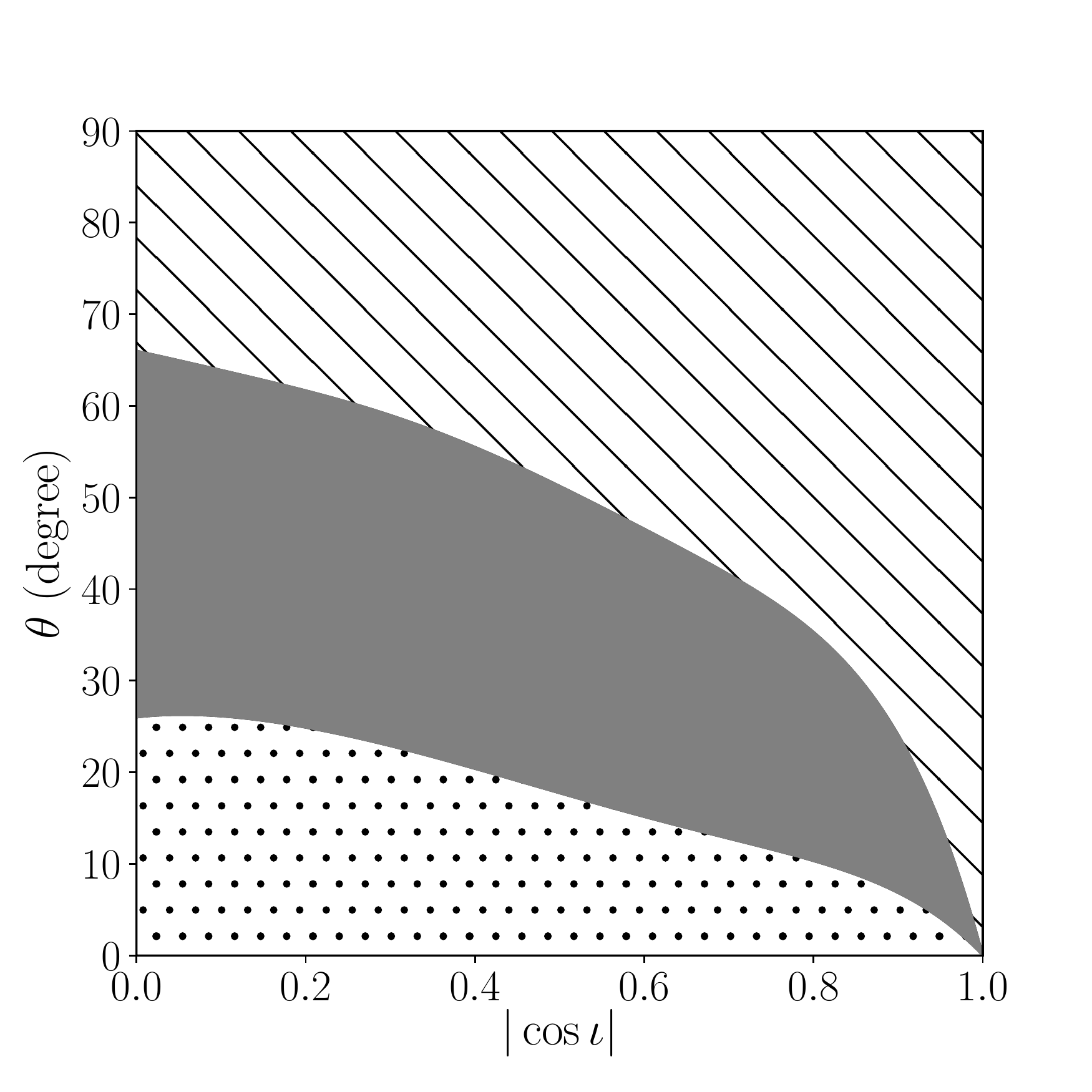}
	\caption[]{Optimal choice of methods as a function of $|\cos \iota|$ and $\theta$. The regions marked by lines, solid gray color, and dots indicate the parameter space where the best sensitivity can be obtained by single component tracking ($2\fstar$ dominates), dual-harmonic tracking, and single component tracking ($\fstar$ dominates), respectively.}
	\label{fig:area}
\end{figure}

We carry out a third set of simulations to determine the optimal tracking method over the whole $(\theta, |\cos \iota|)$ plane. For each $|\cos \iota| \in \{0,0.1,0.2,0.3,0.4,0.5,0.6,0.7,0.8,0.9,1\}$, we run Monte Carlo simulations by injecting signals with various $h_0$ and $\theta$ values. The other parameters and configurations are kept the same as the first and second sets. For each choice of $|\cos \iota|$, we find out two $\theta$ values when $h_0$ is near the detection limit: one yields the same detection efficiency between the $2\fstar$ tracking and dual-harmonic tracking; the other yields the same detection efficiency between the $\fstar$ tracking and dual-harmonic tracking. By connecting these resulting $(|\cos \iota|, \theta)$ points, the two curves correspond to two boundaries: (1) between where the $2\fstar$ component dominates and where both $\fstar$ and $2\fstar$ components contribute, and (2) between where both $\fstar$ and $2\fstar$ components contribute and where the $\fstar$ component dominates. The results are shown in Figure~\ref{fig:area}. The regions marked by lines, solid gray color, and dots indicate the parameter space where the optimal method is single component tracking ($2\fstar$ dominates), dual-harmonic tracking, and single component tracking ($\fstar$ dominates), respectively. 
Generally speaking, for about $1/3$ of the whole $(|\cos \iota|,\theta)$ parameter space (gray region), dual-harmonic tracking performs much better than single frequency tracking, improving detection efficiency by up to $\sim 70\%$. 

\section{Discussion}
\label{sec:discussion}

In this section, we discuss the computing cost of dual-harmonic HMM tracking, and the justifications of applying it to upcoming directed continuous-wave searches. 
Without prior knowledge of the intrinsic parameters of the source, which determine if the gravitational-wave emission is dominated by a single frequency component, or the combination of both components, the optimal sensitivity can be obtained over the whole ($|\cos \iota|$, $\theta$) parameter space by conducting both the conventional single component tracking and the dual-harmonic tracking.
Here we quantify the computing cost of conducting both ways of tracking in a directed search.

The Viterbi algorithm uses dynamic programming \footnote{Dynamic programming, a technique based on Bellman's Principle of Optimality, is used to solve an optimization problem by breaking down the problem into sub-problems of optimization, and making intermediate decisions for sub-problems to reconstruct the final decision in a recursive manner \cite{Bellman1954,Bellman1957,Bertsekas2005}. A detailed description is given in Ref.~\cite{Suvorova2016}.} and reduces the total number of comparisons required to calculate $Q^*(O)$ from $N_Q^{N_T+1}$ to $(N_T+1)N_Q^2$ \cite{Quinn2001,Suvorova2016}. As an example, if we take $N_Q = 10^6$ frequency bins, $N_T= 50$ tracking steps, and $A_{q_j q_i}$ with only three nonzero terms along the diagonal, the total number of comparison is reduced from $10^{30}$ to $10^{8}$.
The cost of computing $Q^*(O)$ (e.g., $\lesssim 1$\,min) is generally negligible compared to that of computing $\mathcal{F}$-statistic values over $N_T$ blocks of $\Tcoh$ (e.g., $\sim 1$\,hr), in a sub-band.
Hence the computing time of a conventional single component tracking over \mbox{$\Tobs = N_T \Tcoh$} in a frequency band from $f_{\rm min}$ to $f_{\rm max}$ is mainly dominated by calculating $N_T$ blocks of $\mathcal{F}$-statistic, given by \cite{Sun2018}
\begin{equation}
\label{eqn:compute_time}
\mathcal{T} = 10\,{\rm d}\left(\frac{f_{\rm max}-f_{\rm min}}{1\,{\rm kHz}}\right) \left(\frac{T_{\rm coh}}{5\,{\rm d}}\right)^2 \left(\frac{N_T}{50}\right) \left(\frac{1}{N_{\rm core}}\right),
\end{equation}
where $N_{\rm core}$ is the number of cores running in parallel.
Given the $N_T$ blocks of $\mathcal{F}$-statistic calculated already for the full frequency band, conducting a dual-harmonic tracking using the same set of $\mathcal{F}$-statistic data barely introduces additional cost, i.e., the total computing time of conducting both ways of tracking can be approximated by Eqn.~\eqref{eqn:compute_time}, yielding a typical run time of $\sim 10^3$\,core-hr for one year's observation.


In addition to improving search sensitivity, dual-harmonic HMM tracking can be used as a candidate follow-up tool in both directed and all-sky continuous-wave searches. When we have a list of above-threshold candidates for further scrutiny as the output from existing directed or all-sky search methods, we can conduct a follow-up procedure as follows. For each candidate at frequency $f_0$, we conduct (a) a HMM tracking in a narrow band around $f_0$ only, and (b) a dual-harmonic HMM tracking in narrow bands around frequencies (1) $f_0$ and $2f_0$, and (2) $f_0$ and $0.5f_0$ (because we have no knowledge if $f_0$ is corresponding to $\fstar$ or $2\fstar$). 
Seeing a more significant detection statistic in (b) than (a) increases the probability of a true dual-harmonic astrophysical signal.


One interesting question is how likely there is a $\fstar$ component in the signal with an amplitude that could benefit the search by taking it into consideration. In other words, is it physically likely that the source parameters lie in the gray region in Figure~\ref{fig:area}? For a freely precessing star, the wobble angle is believed to damp (for oblate deformations) or increase towards $\pi/2$ (for prolate deformations) on an internal dissipation timescale \cite{Alpar1988,Jones2002,Cutler2001,Cutler2002}, making a dual-harmonic search less interesting. However, in the model proposed by Ref.~\cite{Jones2010}, the nonprecessing solution indicates that the star's rotation axis lies closely to the superfluid pinning axis, allowing $0\leq \theta \leq \pi/2$.\footnote{We do not discuss the full range $0\leq \theta \leq \pi$ in Sec.~\ref{sec:simulation}, because $\Phi_0$ and $\theta$ are degenerate. We have $0\leq \theta \leq \pi/2$ and $0\leq \Phi_0 \leq 2\pi$, or $0\leq \theta \leq \pi$ and $0\leq \Phi_0 \leq \pi$ \cite{Jones2015}.} It motives conducting dual-harmonic HMM tracking in future directed searches or candidate follow-ups. More interestingly, detecting or confirming a signal using this method would provide important information for probing the neutron star structure and emission mechanism, e.g., a pinned superfluid interior.

\section{Conclusion}
\label{sec:conclusion}

In this paper, we describe an economical dual-harmonic tracking scheme based on a HMM and combined with the coherent $\mathcal{F}$-statistic, which provides a semicoherent search strategy taking into consideration a model that gravitational-wave emission from a neutron star is at both $\fstar$ and $2\fstar$. We review the signal waveforms and frequency domain estimator, formulate the problem with an extended HMM, discuss the performance analytically based on the distribution of path probabilities, and demonstrate the advantages of the method through Monte Carlo simulations. 

We find that for sources emitting at both $\fstar$ and $2\fstar$, we can improve the detection efficiency by $\sim 10\%$--70\% for $20^\circ \lesssim \theta \lesssim 60^\circ$ by tracking both frequencies simultaneously, compared to a conventional single component search. While at low SNR, dual-harmonic tracking can lead to minor sensitivity loss, reducing detection efficiency by $\lesssim 10\%$, if the source emits at $2\fstar$ only. To achieve the optimal sensitivity in a directed search, we can add the dual-harmonic tracking as a complementary procedure to the conventional single frequency tracking in the full band. The economical HMM tracking algorithm allows conducting both the dual-harmonic tracking and the conventional search at almost no additional cost. 

The method also serves as a useful candidate follow-up tool in the near future when more candidates will be considered for further scrutiny in directed or all-sky continuous-wave searches. Upon detection, the resulting statistics from dual-harmonic tracking and single frequency tracking can shed light on the structure and emission mechanism of a neutron star. In addition, when a better understood model is available in the future for a postmerger remnant from a binary neutron star coalescence, we can apply a similar dual-harmonic tracking scheme to improve the sensitivity in searches for signals from the remnant, considering the possibility that the remnant is freely precessing. 

\section{Acknowledgments}

We are grateful to the LIGO and Virgo Continuous Wave Working Group for informative discussions, and S. Walsh for the review and comments.
L. Sun is a member of the LIGO Laboratory.
LIGO was constructed by the California Institute of Technology and Massachusetts Institute of Technology with funding from the National Science Foundation, and operates under cooperative agreement PHY--0757058. Advanced LIGO was built under award PHY--0823459. 
P. D. Lasky is supported through ARC Future Fellowship FT160100112 and Discovery Project DP180103155.
The research is also supported by Australian Research Council (ARC) Discovery Project DP170103625 and the ARC Centre of Excellence for Gravitational Wave Discovery CE170100004. 
This paper carries LIGO Document Number \dcc.

\end{document}